\documentclass[eqsecnum,showpacs,preprint,amsmath,aps,nofootinbib]{revtex4}  %
\usepackage{epsfig}
\usepackage{amssymb}
\usepackage[dvips]{color}
\usepackage[latin1]{inputenc}
\usepackage{graphicx}

\def\beq{\begin{equation}}
\def\eeq{\end{equation}}
\def\beqq{\begin{eqnarray}}
\def\eeqq{\end{eqnarray}}

\newcommand{\bdm}{\begin{displaymath}}
\newcommand{\edm}{\end{displaymath}}

\def\pmb#1{\setbox0=\hbox{$#1$}%
  \kern-.025em\copy0\kern-\wd0
  \kern.05em\copy0\kern-\wd0
  \kern-.025em\raise.0433em\box0}

\makeatletter
\renewcommand*{\@fnsymbol}[1]{\ensuremath{\ifcase#1\or *\or \dagger\or
    \ddagger\or 
   \mathsection\or **\or \dagger\dagger
   \or \ddagger\ddagger \else\@ctrerr\fi}}
\makeatother

\begin{document}
\title{Quantum effects on Lagrangian points and displaced periodic orbits  
in the Earth-Moon system}

\author{Emmanuele Battista}
\email[E-mail: ]{ebattista@na.infn.it}
\affiliation{Dipartimento di Fisica, Complesso Universitario 
di Monte S. Angelo, Via Cintia Edificio 6, 80126 Napoli, Italy\\
Istituto Nazionale di Fisica Nucleare, Sezione di Napoli, Complesso Universitario di Monte
S. Angelo, Via Cintia Edificio 6, 80126 Napoli, Italy}

\author{Simone Dell'Agnello}
\email[E-mail: ]{simone.dellagnello@lnf.infn.it}
\affiliation{Istituto Nazionale di Fisica Nucleare, Laboratori Nazionali di Frascati,
00044 Frascati, Italy}

\author{Giampiero Esposito}
\email[E-mail: ]{gesposit@na.infn.it}
\affiliation{Istituto Nazionale di Fisica Nucleare, Sezione di
Napoli, Complesso Universitario di Monte S. Angelo, 
Via Cintia Edificio 6, 80126 Napoli, Italy}

\author{Jules Simo}
\email[E-mail: ]{jules.simo@strath.ac.uk}
\affiliation{Department of Mechanical and Aerospace Engineering, University of Strathclyde, Glasgow, G1 1XJ, United Kingdom}

\date{\today}

\begin{abstract}
Recent work in the literature has shown that the one-loop long distance quantum corrections to the Newtonian
potential imply tiny but observable effects in the restricted three-body problem of celestial mechanics, i.e.,
at the Lagrangian libration points of stable equilibrium the planetoid is not exactly at equal distance from the
two bodies of large mass, but the Newtonian values of its coordinates are changed by a few millimeters 
in the Earth-Moon system. First, we assess such a theoretical 
calculation by exploiting the full theory of the quintic equation, i.e., its reduction to
Bring-Jerrard form and the resulting expression of roots in terms of generalized hypergeometric functions.
By performing the numerical analysis of the exact formulas for the roots, we confirm and slightly improve
the theoretical evaluation of quantum corrected coordinates of Lagrangian libration points of stable equilibrium.
Second, we prove in detail that also for collinear Lagrangian points the quantum corrections are of the
same order of magnitude in the Earth-Moon system.
Third, we discuss the prospects to measure, with the help of laser ranging, the above departure from the
equilateral triangle picture, which is a challenging task. 
On the other hand, a modern version of the planetoid is the solar sail, and much progress
has been made, in recent years, on the displaced periodic orbits of solar sails at all libration points, both
stable and unstable. The present paper investigates therefore, eventually, a restricted three-body problem involving 
Earth, Moon and a solar sail. By taking into account the one-loop
quantum corrections to the Newtonian potential, displaced 
periodic orbits of the solar sail at libration points are again found to exist.
\end{abstract}


\maketitle

\section{Introduction}

From the point of view of modern theoretical physics, the logical need for a quantum theory of gravity is
suggested by the Einstein equations themselves, which tell us that gravity couples to $T_{\mu \nu}$, the
energy-momentum tensor of matter, in a diffeomorphism-invariant way, by virtue of the tensor equations
\cite{Einstein1916,Bruhat09}
\begin{equation}
R_{\mu \nu}-{1\over 2}g_{\mu \nu}R= {8 \pi G \over c^{4}}T_{\mu \nu}.
\label{(1.1)}
\end{equation}
When Einstein arrived at these equations, although he had already understood that the classical Maxwell theory
of electromagnetic phenomena is not valid in all circumstances, the only known forms of $T_{\mu \nu}$ were
classical, e.g., the energy-momentum tensor of a relativistic fluid, or even just the case of vacuum Einstein 
equations, for which $T_{\mu \nu}$ vanishes. In due course, it was realized that matter fields are quantum fields
in the first place (e.g., a massive Dirac field, or spinor electrodynamics). The quantum fields are
operator-valued distributions \cite{Wightman96}, for which a regularization and renormalization procedure is 
necessary and even fruitful. However, the mere replacement of $T_{\mu \nu}$ by its regularized and renormalized
form $\langle T_{\mu \nu} \rangle$ on the right-hand side of Eq. (1.1) leads to a hybrid scheme, because the
classical Einstein tensor $R_{\mu \nu}-{1 \over 2}g_{\mu \nu}R$ is affected by the coupling to
$\langle T_{\mu \nu} \rangle$. The question then arises whether the appropriate, full quantum theory of gravity
should have field-theoretical nature or should involve, instead, other structures 
(e.g., strings \cite{Polchinski98} or loops \cite{Rovelli04} or twistors \cite{Penrose73}), 
and at least $16$ respectable approaches \cite{Espo11} have been developed so far in the literature. To make such
theories truly physical, their predictions should be checked against observations. For example, applications of
the covariant theory lead to detailed predictions for the cross sections of various scattering processes
\cite{DeWitt1967c}, but such phenomena (if any) occur at energy scales inaccessible to observations, and also
the effects of Planck-scale physics on cosmology, e.g., the cosmic microwave background radiation and its 
anisotropy spectrum \cite{Kiefer12,Bini13,Bini14,Kamen13,Kamen14}, are not yet easily accessible to observations, although 
cosmology offers possibly the best chances for testing quantum gravity \cite{Wilczek14}.

Recently, inspired by the work in Refs. \cite{D94,D94b,D94c,MV95,HL95,ABS97,KK02,D03} on effective field theories of
gravity, where it is shown that the leading (i.e., one-loop) long distance quantum corrections to the Newtonian potential are
entirely ruled by the Einstein-Hilbert part of the full action functional, some of us 
\cite{BE14a,BE14b} have assumed that such a 
theoretical analysis can be applied to the long distances and macroscopic bodies occurring in celestial    
mechanics \cite{P1890,P1892,Pars65}. More precisely, the Newtonian potential between two bodies of masses
$m_{A}$ and $m_{B}$ receives quantum corrections leading to \cite{D03}
\begin{equation}
V_{Q}(r)=-{G m_{A}m_{B}\over r} \left(1+{k_{1}\over r}+{k_{2}\over r^{2}} \right) 
+{\rm O}(G^{2}),
\label{(1.2)}
\end{equation}
where \cite{BE14a}
\begin{equation}
k_{1} \equiv \kappa_{1}{G (m_{A}+m_{B})\over c^{2}},
\label{(1.3)}
\end{equation}
\begin{equation}
k_{2} \equiv \kappa_{2}{G \hbar \over c^{3}}= \kappa_{2}(l_{P})^{2}.
\label{(1.4)}
\end{equation}
Equation (1.2) implies that, $\forall \varepsilon >0$, there exists an $r_{0}$ value of $r$
such that
\begin{equation}
\left | V_{Q}(r) + {G m_{A}m_{B}\over r}\left(1+{k_{1}\over r}
+{k_{2}\over r^{2}}\right)\right | < \varepsilon, \; \forall r > r_{0}.
\label{(1.5)}
\end{equation}
This feature will play an important role in our concluding remarks in Sec. V.

We also stress that the dimensionless parameter $\kappa_{1}$ depends on the dimensionless parameter
$\kappa_{2}$. In other words, $k_{1}$ is a post-Newtonian term which only depends on classical physical
constants, but its weight, expressed by the real number $\kappa_{1}$, is affected by the calculational
procedure leading to the fully quantum term $k_{2}$, where the real number $\kappa_{2}$ weighs the
Planck length squared. More precisely, the perturbative expansion involves only integer powers of
Newton's constant G:
\begin{equation}
V_{Q}(r) \sim -{G m_{A}m_{B}\over r} \left(1+\sum_{p=1}^{\infty}f_{p}(r)G^{p} \right)
\sim -{G m_{A}m_{B}\over r} \left(1+\sum_{n=1}^{\infty}{k_{n}\over r^{n}}\right),
\label{(1.6)}
\end{equation}
where, upon denoting by $L_{A}$ and $L_{B}$ the gravitational radii
$L_{A} \equiv {G m_{A}\over c^{2}},L_{B} \equiv {G m_{B}\over c^{2}}$, one has
the coefficients $k_{n}=k_{n}(L_{A}+L_{B},(l_{P})^{2})$. At one loop, i.e., to linear order in
$G$, where
\begin{equation}
f_{1}(r)=\kappa_{1}{(m_{A}+m_{B})\over c^{2}}{1 \over r}
+\kappa_{2}{\hbar \over c^{3}} {1 \over r^{2}},
\label{(1.7)}
\end{equation}
we can only have the contribution ${(L_{A}+L_{B})\over r}$ with weight equal to the real number
$\kappa_{1}$, and the contribution ${(l_{P})^{2}\over r^{2}}$ with weight equal to the real
number $\kappa_{2}$. Although the term ${(l_{P})^{2}\over r^{2}}$ is overwhelmed by the term
${(L_{A}+L_{B})\over r}$, the two are inextricably intertwined because $\kappa_{1}$ is not a free
real parameter but depends on $\kappa_{2}$: both $\kappa_{1}$ and $\kappa_{2}$ result from loop
diagrams. Thus, the one-loop long distance quantum correction {\it is the whole term}
\begin{equation}
f_{1}(r)G={k_{1}\over r}+{k_{2}\over r^{2}}=\kappa_{1}{(L_{A}+L_{B})\over r}
+\kappa_{2}{(l_{P})^{2}\over r^{2}},
\label{(1.8)}
\end{equation}
where $\kappa_{1}$ takes a certain value because there exists a nonvanishing value of $\kappa_{2}$.
The authors of Ref. \cite{D03} found the numerical values
\begin{equation}
\kappa_{1}=3, \; \kappa_{2}={41 \over 10 \pi}.
\label{(1.9)}
\end{equation}
The work in Ref. \cite{BE14a} considered the application of Eqs. (1.2)-(1.4) 
and (1.9) to the circular restricted
three-body problem of celestial mechanics, in which two bodies $A$ and $B$ of masses $\alpha$ and $\beta$,
respectively, with $\alpha > \beta$, move in such a way that the orbit of $B$ relative to $A$ is circular,
and hence both $A$ and $B$ move along circular orbits around their center of mass 
$C$ which moves in a straight line or is at rest, while a third body,
the planetoid $P$, of mass $m$ much smaller than $\alpha$ and $\beta$, is subject to their gravitational
attraction, and one wants to evaluate the motion of the planetoid. On taking 
rotating axes\footnote{To be self-consistent, some minor repetition of the text in Ref. 
\cite{BE14a} is unavoidable.} with center
of mass $C$ as origin, distance $AB$ denoted by $l$, and angular velocity $\omega$ given by
\begin{equation}
\omega=\sqrt{G(\alpha+\beta)\over l^{3}},
\label{(1.10)}
\end{equation}
one has, with the notation in Fig. 1, that the quantum corrected effective potential for the circular
restricted three-body problem is given by $GU$, where \cite{BE14a}
\begin{equation}
U={1\over 2}{(\alpha+\beta)\over l^{3}}(x^{2}+y^{2})
+{\alpha \over r} \left(1+{k_{1}\over r}+{k_{2}\over r^{2}}\right)
+{\beta \over s} \left(1+{k_{3}\over s}+{k_{2}\over s^{2}}\right),
\label{(1.11)}
\end{equation}
where $r$ and $s$ are the distances $AP$ and $BP$, respectively, while here
\begin{equation}
k_{1} \equiv \kappa_{1}{G (m+\alpha)\over c^{2}}, \;
k_{3} \equiv \kappa_{3}{G (m+\beta)\over c^{2}},
\kappa_{1}=\kappa_{3}=3.
\label{(1.12)}
\end{equation}
The equilibrium points are found by studying the gradient of $U$ and evaluating its zeros. There exist
indeed five zeros of ${\rm grad}U$ \cite{BE14a}. Three of them correspond to collinear libration points
$L_{1},L_{2},L_{3}$ of unstable equilibrium, while the remaining two describe configurations of stable
equilibrium at the points denoted by $L_{4},L_{5}$. The simple but nontrivial idea in Refs. 
\cite{BE14a,BE14b} was that, even though the quantum corrections in (1.2) involve small quantities, 
{\it the analysis of stable equilibrium} (to linear order in perturbations) 
{\it might lead to testable departures from Newtonian theory, being related to the gradient of} 
$U$, and to the second derivatives of $U$ evaluated
at the zeros of ${\rm grad}U$. The quantum corrected Lagrange points $L_{4}$ and $L_{5}$ have 
coordinates $(x(l),y_{+}(l))$ and $(x(l),y_{-}(l))$, respectively, where
\begin{equation}
x(l)={(r^{2}(l)-s^{2}(l)+b^{2}-a^{2})\over 2(a+b)},
\label{(1.13)}
\end{equation}
\begin{equation}
y_{\pm}(l)= \pm \sqrt{r^{2}(l)-x^{2}(l)-2a x(l)-a^{2}},
\label{(1.14)}
\end{equation}
where $r(l) \equiv {1 \over w(l)}, \; s(l) \equiv {1 \over u(l)}$, $w(l)$ ans $u(l)$ being the real solutions of 
an algebraic equation of fifth degree (see Sec. II). Interestingly, $r(l) \not = s(l)$, and hence to the 
equilateral libration points of Newtonian celestial mechanics 
there correspond points no longer exactly at vertices of an 
equilateral triangle. For the Earth-Moon-satellite system, the work in Ref. \cite{BE14b} has found
\begin{equation}
x_{Q}-x_{C} \approx 8.8 \; {\rm mm}, \; |y_{Q}|-|y_{C}| \approx 5.2 \; {\rm mm},
\label{(1.15)}
\end{equation}
where $x_{Q}$ (resp. $x_{C}$) is the quantum corrected (resp. classical) value of $x(l)$ in (1.13),
and the same for $y_{Q}$ and $y_{C}$ obtainable from (1.14). Remarkably, the values in (1.15) are well
accessible to the modern astrometric techniques \cite{Pitjeva09}.

\begin{figure}
\includegraphics[scale=0.70]{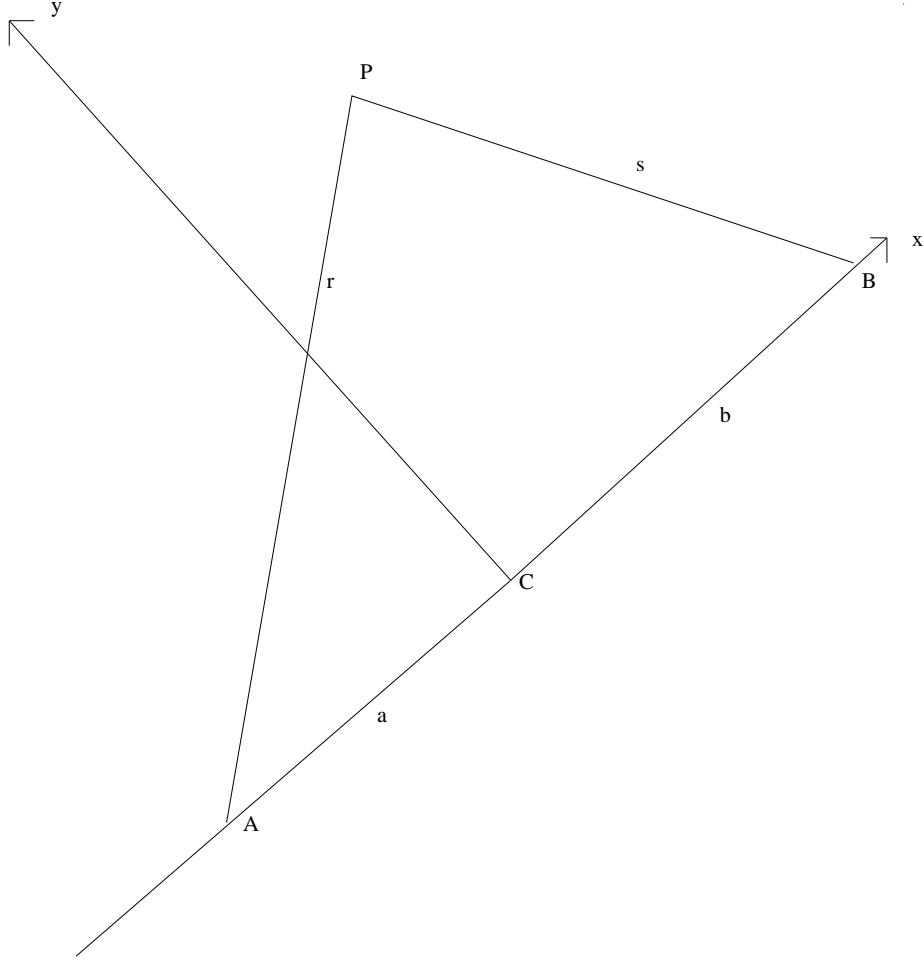}
\caption{The figure shows the two bodies of large mass,
$A$ and $B$, the center of mass $C$, and the planetoid at $P$.}
\end{figure}

On the other hand, much progress has been made along the years on modern models of planetoids and their
displaced periodic orbits at all Lagrange points $L_{1},L_{2},L_{3},L_{4},L_{5}$, to linear order in
the variational equations for Newtonian theory. In particular, a modern version of planetoid is a solar
sail, which is propelled by reflecting solar photons and therefore can transform the momentum of photons
into a propulsive force. Solar sailing technology appears as a promising form of advanced spacecraft
propulsion \cite{Simo08,Simo09a,Simo09b,Simo10,Simo14}, 
which can enable exciting new space-science mission concepts such as solar
system exploration and deep space observation. Although solar sailing has been considered as a practical
means of spacecraft propulsion only relatively recently, the fundamental ideas had been already developed
towards the end of the previous century \cite{McInnes1999}.

Solar sails can also be used for highly nonKeplerian orbits, such as closed orbits displaced high above
the ecliptic plane \cite{Waters07}. Solar sails are especially suited for such nonKeplerian orbits, since
they can apply a propulsive force continuously. This makes it possible to consider some exciting and unique
trajectories. In such trajectories, a sail can be used as a communication satellite for high latitudes. For
example, the orbital plane of the sail can be displaced above the orbital plane of the Earth, so that the sail
can stay fixed above the Earth at some distance, if the orbital periods are equal. Orbits around the collinear
points of the Earth-Moon system are also of great interest because their unique positions are advantageous
for several important applications in space mission design 
\cite{Szebehely67,Roy05,Vonbun68,Thurman96,Gomez01a,Gomez01b}.

Over the last few dacades, several authors have tried to determine more accurate approximations of such equilibrium
points \cite{Farquhar73}. Such (quasi-)Halo orbits were first studied in Refs. 
\cite{Farquhar73,Farquhar71,Breakwell79,Richardson80,Howell84,Howell05}. Halo orbits near the collinear libration
points in the Earth-Moon system are of great interest, in particular around the $L_{1}$ and $L_{2}$ points,
because of their unique positions. However, a linear analysis shows that the collinear libration points
$L_{1},L_{2}$ and $L_{3}$ are of the type saddle$\times$center$\times$center, leading to an instability in
their vicinity, whereas the equilateral equilibrium points $L_{4}$ and $L_{5}$ are stable, in that they are
of the type center$\times$center$\times$center. Although the libration points $L_{4}$ and $L_{5}$ are naturally
stable and require a small acceleration, the disadvantage is the longer communication path length from the
lunar pole to the sail. 
If the orbit maintains visibility from Earth, a spacecraft on it (near the $L_{2}$ point) can be used to provide
communications between the equatorial regions of the Earth and the polar regions of the Moon. The establishment of
a bridge for radio communications is crucial for forthcoming space missions, which plan to use the lunar poles.
Displaced nonKeplerian orbits near the Earth-Moon libration points have been investigated in Refs. 
\cite{McInnes1993,Simo08,Simo09a,Simo09b,Simo10,Simo14}. This brief outline shows therefore that the analysis 
of libration points does not belong just to the history of celestial mechanics, but plays a crucial 
role in modern investigations of space mission design.  

Section II studies in detail the algebraic equation of fifth degree for the evaluation of noncollinear libration
points $L_{4}$ and $L_{5}$, by first passing to dimensionless units and then exploiting the rich mathematical
theory of quintic equations and their roots. Section III derives and solves the algebraic equation of ninth
degree for the evaluation of quantum corrections to collinear Lagrangian points $L_{1},L_{2},L_{3}$.
Section IV outlines the prospects to measure the quantum 
corrected coordinates obtained in Sec. II with the help of laser ranging.
Section V evaluates displaced periodic orbits at the 
quantum-corrected Lagrange points $L_{4}$ and $L_{5}$, and a detailed comparison with the results of Newtonian
celestial mechanics is also made. Concluding remarks and open problems are presented in Sec. VI, while the
Appendices describe relevant background material on the theory of algebraic equations.

\section{Algebraic equations of fifth degree for $w(l)$ and $u(l)$}

In Ref. \cite{BE14a} it has been shown that the ${\rm grad}U=0$ condition at noncollinear libration points leads
to the algebraic equations of fifth degree
\begin{equation}
w^{5}+\zeta_{4}w^{4}+\zeta_{3}w^{3}+\zeta_{0}=0,
\label{(2.1)}
\end{equation}
\begin{equation}
u^{5}+{\tilde \zeta}_{4}u^{4}+{\tilde \zeta}_{3}u^{3}+{\tilde \zeta}_{0}=0,
\label{(2.2)}
\end{equation}
where
\begin{equation}
\zeta_{4}={2 \over 3}{\kappa_{1}\over \kappa_{2}}
{G(m+\alpha)\over c^{2}l_{P}^{2}}, \;
\zeta_{3}={1 \over 3 \kappa_{2}}{1 \over l_{P}^{2}}, \;
\zeta_{0}=-{1 \over 3 \kappa_{2}}{1 \over l_{P}^{2}l^{3}},
\label{(2.3)}
\end{equation}
\begin{equation}
{\tilde \zeta}_{4}={2 \over 3}{\kappa_{1}\over \kappa_{2}}{G(m+\beta)\over c^{2}l_{P}^{2}}, \;
{\tilde \zeta}_{3}=\zeta_{3}, \;
{\tilde \zeta}_{0}=\zeta_{0}.
\label{(2.4)}
\end{equation}
Such formulas tell us that it is enough to focus on Eq. (2.1), say, where, to exploit the mathematical theory
of quintic equations, we pass to dimensionless units by defining
\begin{equation}
w={1 \over r} \equiv {\gamma \over l_{P}},
\label{(2.5)}
\end{equation}
where $\gamma$ is a real number to be determined. The quintic equation obeyed by $\gamma$ is therefore
\begin{equation}
\gamma^{5}+\rho_{4}\gamma^{4}+\rho_{3}\gamma^{3}+\rho_{0}=0,
\label{(2.6)}
\end{equation}
where $\rho_{4},\rho_{3},\rho_{0}$ are all dimensionless and read as
\begin{equation}
\rho_{4} \equiv \zeta_{4}l_{P}={2 \over 3}{\kappa_{1}\over \kappa_{2}}
{G(m+\alpha)\over c^{2}l_{P}},
\label{(2.7)}
\end{equation}
\begin{equation}
\rho_{3} \equiv \zeta_{3}l_{P}^{2}={1 \over 3 \kappa_{2}},
\label{(2.8)}
\end{equation}
\begin{equation}
\rho_{0} \equiv \zeta_{0}l_{P}^{5}=-{1 \over 3 \kappa_{2}}
\left({l_{P}\over l}\right)^{3}
=-\rho_{3} \left({l_{P}\over l}\right)^{3}.
\label{(2.9)}
\end{equation}
At this stage, we can exploit the results of Appendix A, by virtue of which Eq. (2.6) can be brought
into the Bring-Jerrard \cite{Bring,Jerrard} form of quintic equations
\begin{equation}
\gamma^{5}+d_{1}\gamma+d_{0}=0.
\label{(2.10)}
\end{equation}
Since we are going to need the roots of the quintic up to the ninth or tenth decimal digit, the form 
(2.10) of the quintic will turn out to be very useful, because it leads to exact formulas for the
roots which are then evaluated numerically, which is possibly better than solving numerically the
quintic from the beginning. Hermite \cite{Hermite} proved that this 
equation can be solved in terms of elliptic functions, but we 
use the even more manageable formulas for the roots displayed in Ref. \cite{Birkeland1924}. For this purpose,
the crucial role is played by the number
\begin{equation}
\sigma \equiv {3125 \over 256}{(-d_{0})^{4}\over (-d_{1})^{5}}.
\label{(2.11)}
\end{equation}
We can further simplify Eq. (2.10) by rescaling $\gamma$ according to
\begin{equation}
\gamma=\chi {\tilde \gamma}.
\label{(2.12)}
\end{equation}
The quintic for ${\tilde \gamma}$ is then
\begin{equation}
{\tilde \gamma}^{5}+{d_{1}\over \chi^{4}}{\tilde \gamma}+{d_{0}\over \chi^{5}}=0. 
\label{(2.13)}
\end{equation}
One can choose $\chi$ in such a way that
\begin{equation}
-{d_{1}\over \chi^{4}}=1 \Longrightarrow \chi=\chi(d_{1})=(-d_{1})^{1 \over 4},
\label{(2.14)}
\end{equation}
and the corresponding $\sigma$ of (2.11) reads as
\begin{equation}
{\tilde \sigma}={3125 \over 256}\left(-{d_{0}\over (\chi(d_{1}))^{5}}\right)^{4}=\sigma.
\label{(2.15)}
\end{equation}
If $|{\tilde \sigma}|<1$, which is the case that holds in the Earth-Moon system by virtue of the numerical
results for $d_{0}$ and $d_{1}$ obtained from the algorithm of Appendix A, then the analysis of Ref.
\cite{Birkeland1924} shows that the five roots of the quintic (2.13), here written as
${\tilde \gamma}^{5}-{\tilde \gamma}-{\tilde \beta}=0$, are obtained from the parameter
\begin{equation}
{\tilde \beta} \equiv -{d_{0}\over (\chi(d_{1}))^{5}}
\label{(2.16)}
\end{equation}
occurring in ${\tilde \sigma}$, and from hypergeometric functions of order $4$, 
according to \cite{Birkeland1924}
\begin{equation}
\left(\begin{matrix}
{\tilde \gamma}_{1} \cr {\tilde \gamma}_{2} \cr {\tilde \gamma}_{3} \cr 
{\tilde \gamma}_{4} 
\end{matrix}\right)
= \left(\begin{matrix}
{\rm i} & {{\tilde \beta}\over 4} & {5 \over 32}{\rm i}{\tilde \beta}^{2} 
& -{5 \over 32}{\tilde \beta}^{3} \cr
-1 & {{\tilde \beta}\over 4} & {5 \over 32}{\tilde \beta}^{2} 
& {5 \over 32}{\tilde \beta}^{3} \cr
-{\rm i} & {{\tilde \beta}\over 4} & -{5 \over 32}{\rm i}{\tilde \beta}^{2} 
& -{5 \over 32}{\tilde \beta}^{3} \cr
1 & {{\tilde \beta}\over 4} & -{5 \over 32}{\tilde \beta}^{2} 
& {5 \over 32}{\tilde \beta}^{3} 
\end{matrix}\right)
\left(\begin{matrix}
F_{0}({\tilde \sigma}) \cr F_{1}({\tilde \sigma}) \cr
F_{2}({\tilde \sigma}) \cr F_{3}({\tilde \sigma})
\end{matrix}\right),
\label{(2.17)}
\end{equation}
\begin{equation}
{\tilde \gamma}_{5}=-{\tilde \beta}F_{1}({\tilde \sigma}),
\label{(2.18)}
\end{equation}
where, having defined the higher hypergeometric function
\begin{equation}
F:{\tilde \sigma} \rightarrow F({\tilde \sigma}) \equiv F \left(\begin{matrix}
a_{1}, & a_{2}, & ..., & a_{n-2}, & a_{n-1} \cr
b_{1}, & b_{2}, & ..., & b_{n-2}, & {\tilde \sigma} 
\end{matrix}\right)
=\sum_{s=0}^{\infty}C_{s}{\tilde \sigma}^{s},
\label{(2.19)}
\end{equation}
with the coefficients evaluated according to the rules
\begin{equation}
C_{0} \equiv 1, \; C_{s} \equiv 
{(a_{1},s)(a_{2},s)...(a_{n-2},s)(a_{n-1},s) \over (1,s)(b_{1},s)...(b_{n-3},s)(b_{n-2},s)},
\label{(2.20)}
\end{equation}
\begin{equation}
(\lambda,\mu) \equiv \lambda (\lambda+1)(\lambda+2)...(\lambda+\mu-1),
\label{(2.21)}
\end{equation}
one has
\begin{equation}
F_{0}({\tilde \sigma}) \equiv F \left(\begin{matrix}
-{1 \over 20}, & {3 \over 20}, & {7 \over 20}, & {11 \over 20} \cr
{1 \over 4}, & {1 \over 2}, & {3 \over 4}, & {\tilde \sigma}
\end{matrix}\right),
\label{(2.22)}
\end{equation}
\begin{equation}
F_{1}({\tilde \sigma}) \equiv F \left(\begin{matrix}
{1 \over 5}, & {2 \over 5}, & {3 \over 5}, & {4 \over 5} \cr
{1 \over 2}, & {3 \over 4}, & {5 \over 4}, & {\tilde \sigma}
\end{matrix}\right),
\label{(2.23)}
\end{equation}
\begin{equation}
F_{2}({\tilde \sigma}) \equiv F \left(\begin{matrix}
{9 \over 20}, & {13 \over 20}, & {17 \over 20}, & {21 \over 20} \cr
{3 \over 4}, & {5 \over 4}, & {3 \over 2}, & {\tilde \sigma}
\end{matrix}\right),
\label{(2.24)}
\end{equation}
\begin{equation}
F_{3}({\tilde \sigma}) \equiv F \left(\begin{matrix}
{7 \over 10}, & {9 \over 10}, & {11 \over 10}, & {13 \over 10} \cr
{5 \over 4}, & {3 \over 2}, & {7 \over 4}, & {\tilde \sigma}
\end{matrix}\right).
\label{(2.25)}
\end{equation}
This representation of the roots ${\tilde \gamma}_{i}$ is discovered by pointing out that Eqs. (2.13)
and (2.14) suggest considering such roots as functions of ${\tilde \sigma}=\sigma$. By taking 
derivatives of Eq. (2.13) with respect to $\sigma$ up to the fourth order, one can then prove that
all ${\tilde \gamma}_{i}$ are particular integrals of the fourth-order ordinary differential 
equation \cite{Birkeland1924}
\begin{equation}
\left[\sigma^{3}(\sigma-1){{\rm d}^{4}\over {\rm d} \sigma^{4}}
+\sigma^{2}(A_{1}\sigma-B_{1}){{\rm d}^{3}\over {\rm d}\sigma^{3}}
+\sigma(A_{2}\sigma-B_{2}){{\rm d}^{2}\over {\rm d}\sigma^{2}}
+(A_{3}\sigma-B_{3}){{\rm d}\over {\rm d}\sigma}+C \right]\Lambda=0,
\label{(2.26)}
\end{equation}
where $A_{1},A_{2},A_{3},B_{1},B_{2},B_{3},C$ are constants. The roots ${\tilde \gamma}_{i}$ undergo a
peculiar variation when ${\tilde \beta}$ describes an arbitrary curve in its plane. The critical
points turn out to be
\begin{equation}
{\tilde \beta}_{1}=-{\rm i}C, \; {\tilde \beta}_{2}=C, \;
{\tilde \beta}_{3}={\rm i}C, \; {\tilde \beta}_{4}=-C, \; C \equiv {1024 \over 3125}.
\label{(2.27)}
\end{equation}
The group of the linear differential equation (2.26) has in this case the property that
the root ${\tilde \gamma}_{k}$ is changed into ${\tilde \gamma}_{5}$, for all $k=1,2,3,4$,
when ${\tilde \beta}$ describes a small closed contour about the critical point
${\tilde \beta}_{k}$.

Eventually, the roots $\gamma_{i}$ of Eq. (2.10) are given by
\begin{equation}
\gamma_{i}=\chi(d_{1}){\tilde \gamma}_{i}, \;
\forall i=1,2,3,4,5 .
\label{(2.28)}
\end{equation}
At this stage, we have to invert the cubic transformation (A9) 
to find the five roots of the original quintic equation (2.6). Since in this equation the number of
sign differences between consecutive nonvanishing coefficients is $1$, we know from Descartes'
sign rule that it has only one positive root. We find for its numerical value 
(from the definition (2.5) it is clear that only 
positive values of $\gamma$ are physically admissible)
\begin{equation}
\gamma_{+}=4.208852239482695 \cdot 10^{-44}.
\label{(2.29)}
\end{equation}
This value is not affected by the planetoid mass $m$, since $m$ is much smaller than $\alpha$ in (2.7).
As far as the unphysical roots\footnote{We do not need many decimal digits for unphysical roots.} are 
concerned, two of them are real and negative, i.e.
\begin{equation}
\gamma_{-}=-4.20 \cdot 10^{32} \; {\rm or} \; -6.07 \cdot 10^{-34},
\label{(2.30)}
\end{equation}
and two of them are complex conjugate, i.e.
\begin{equation}
\gamma_{C}=-2.10 \cdot 10^{-44} \pm {\rm i} 3.36 \cdot 10^{-44}.
\label{(2.31)}
\end{equation}
Similarly, by repeating the whole analysis for
\begin{equation}
u={1 \over s} \equiv {\Gamma \over l_{P}},
\label{(2.32)}
\end{equation}
we find, by virtue of (2.3) and (2.4), only one positive root
\begin{equation}
\Gamma_{+}=4.208852239579132 \cdot 10^{-44},
\label{(2.33)}
\end{equation}
which is not affected by the planetoid mass $m$, since $m$ is much smaller than $\beta$ in (2.4),
whereas, among the unphysical roots, two are real and negative:
\begin{equation}
\Gamma_{-}=-5.17 \cdot 10^{30} \; {\rm or} \; -4.93 \cdot 10^{-32},
\label{(2.34)}
\end{equation}
while the remaining two are complex conjugate and read as
\begin{equation}
\Gamma_{C}=-2.10 ... \cdot 10^{-44} \pm {\rm i} 3.36 ... \cdot 10^{-44},
\label{(2.35)}
\end{equation}
where the ellipsis denotes that a very tiny difference occurs in the decimal digits with respect
to the result in (2.31), beginning at the eleventh decimal digit for the real part and at the
tenth decimal digit for the imaginary part. 

At this stage, we can exploit Eqs. (1.13) and (1.14) to evaluate the coordinates of quantum corrected
Lagrange points of the Earth-Moon system, finding that
\begin{equation}
x_{Q}=1.8752814881103817 \cdot 10^{8} \; {\rm m}, \;
y_{Q}=\pm 3.329001652107382 \cdot 10^{8} {\rm m}.
\label{(2.36)}
\end{equation}
The Newtonian values of such coordinates are instead
\begin{equation}
x_{C}=1.8752814880224872 \cdot 10^{8} \; {\rm m}, \;
y_{C}=\pm 3.329001652147382 \cdot 10^{8} {\rm m}.
\label{(2.37)}
\end{equation}
Interestingly, our detailed analysis confirms therefore the orders of magnitude
found in Refs. \cite{BE14a,BE14b}, because we obtain (cf. Eq. (1.15))
\begin{equation}
x_{Q}-x_{C} \approx 8.7894 \; {\rm mm}, \;
|y_{Q}|-|y_{C}| \approx -4 \; {\rm mm}.
\label{(2.38)}
\end{equation}
More precisely, our refined analysis confirms to a large extent the theoretical value of
$x_{Q}-x_{C}$, whereas the sign of $|y_{Q}|-|y_{C}|$ gets reversed with respect to Eq. (1.15),
and its magnitude gets reduced by $20$ per cent.

\section{Collinear libration points}

From the theoretical point of view it is equally important to work out how the Lagrangian points 
of unstable equilibrium, usually denoted by $L_{1},L_{2},L_{3}$, get affected by the one-loop
long-distance quantum corrections to the Newtonian potential. On the side of the applications, 
their importance is further strengthened, since satellites 
(e.g., the Wilkinson Microwave Anisotropy Probe) have been sent so far to the 
points $L_{1},L_{2}$ and $L_{3}$ of some approximate three-body configurations in the solar system.

We beging by recalling from Ref. \cite{BE14a} that the gradient of the effective potential in the
restricted three-body problem has components
\begin{eqnarray}
{\partial U \over \partial x}&=& 
{(\alpha+\beta)\over l^{3}}x
-{\alpha x \over r^{3}}
\left(1+2{k_{1}\over r}+3{k_{2}\over r^{2}}\right) 
-{\beta x \over s^{3}}
\left(1+2{k_{3}\over s}+3{k_{2}\over s^{2}}\right) 
\nonumber \\
&+& {\alpha \beta l \over (\alpha+\beta)}
\left[{1\over s^{3}}
\left(1+2{k_{3}\over s}+3{k_{2}\over s^{2}}\right) 
-{1\over r^{3}}
\left(1+2{k_{1}\over r}+3{k_{2}\over r^{2}}\right)\right],
\label{(3.1)}
\end{eqnarray}
\begin{equation}
{\partial U \over \partial y}=y \left[
{(\alpha+\beta)\over l^{3}}
-{\alpha \over r^{3}}
\left(1+2{k_{1}\over r}+3{k_{2}\over r^{2}}\right) 
-{\beta \over s^{3}}
\left(1+2{k_{3}\over s}+3{k_{2}\over s^{2}}\right)\right].
\label{(3.2)}
\end{equation}
When the libration points are collinear, the coordinate $y$ vanishes, which ensures the vanishing of 
${\partial U \over \partial y}$ as well. On the other hand, from the geometry of the problem, as
shown in Fig. 1, one has
\begin{equation}
y^{2}=r^{2}-x^{2}-2ax-a^{2}.
\label{(3.3)}
\end{equation}
The vanishing of $y$ implies therefore that $x$ obeys the algebraic equation
\begin{equation}
x^{2}+2ax+a^{2}-r^{2}=0,
\label{(3.4)}
\end{equation}
which is solved by the two roots
\begin{equation}
x=\varepsilon r -a=\varepsilon r -{\beta l \over (\alpha+\beta)}, \; 
\varepsilon= \pm 1.
\label{(3.5)}
\end{equation}
Furthermore, the geometry of the problem yields also
\begin{equation}
x={(r^{2}-s^{2})\over 2l}+{1 \over 2}{(\alpha-\beta)\over (\alpha+\beta)}l,
\label{(3.6)}
\end{equation}
which implies, by comparison with Eq. (3.5),
\begin{equation}
s^{2}=(r -\varepsilon l)^{2} \Longrightarrow s= \pm (r -\varepsilon l),
\label{(3.7)}
\end{equation}
where both signs should be considered, since $(r-\varepsilon l)$ may be negative.
Note now that the insertion of (3.5) into Eq. (3.1) yields
\begin{equation}
{\partial U \over \partial x}={\beta \over s^{3}}\left(1+2{k_{3}\over s}
+3{k_{2}\over s^{2}}\right)(l- \varepsilon r)
-{\alpha \varepsilon \over r^{2}}\left(1+2{k_{1}\over r}+3{k_{2}\over r^{2}}
\right)+{(\alpha+\beta)\over l^{3}}\left(\varepsilon r -{\beta l \over (\alpha+\beta)}\right)=0.
\label{(3.8)}
\end{equation}
Moreover, we consider first the solution $s=r - \varepsilon l$ in Eq. (3.7). This turns Eq. (3.8)
into the form
\begin{equation}
{\beta \over (r-\varepsilon l)^{2}}+{2k_{3}\beta \over (r-\varepsilon l)^{3}}
+{3 k_{2}\beta \over (r-\varepsilon l)^{4}}
+{\alpha \over r^{2}}+{2 k_{1}\alpha \over r^{3}}+{3 k_{2} \alpha \over r^{4}}
-{(\alpha+\beta)r \over l^{3}}+{\beta \varepsilon \over l^{2}}=0.
\label{(3.9)}
\end{equation}
This form of the equation to be solved for $r={\overline {AP}}$ suggests multiplying 
both sides by $(r-\varepsilon l)^{4}r^{4}$, which makes it clear that we end up by studying a
nonic algebraic equation. Moreover, it is now convenient to adopt dimensionless units. For this
purpose, we point out that the length parameters $k_{1}$ and $k_{3}$ in the potential (1.11) are a linear
combination of the gravitational radii $L_{\alpha},L_{\beta}$ of primaries and $L_{m}$ of planetoid
according to the relations
\begin{equation}
L_{\alpha} \equiv {G \alpha \over c^{2}}, \; L_{\beta} \equiv {G \beta \over c^{2}}, \;
L_{m} \equiv {Gm \over c^{2}},
\label{(3.10)}
\end{equation}
\begin{equation}
l_{\alpha}=L_{\alpha}+L_{m}, \; l_{\beta}=L_{\beta}+L_{m},
\label{(3.11)}
\end{equation}
\begin{equation}
k_{1}=\kappa_{1} l_{\alpha}, \; k_{3}=\kappa_{3}l_{\beta}=\kappa_{1}l_{\beta},
\label{(3.12)}
\end{equation}
while $k_{2}=\kappa_{2}(l_{P})^{2}$ from Eq. (1.4). Furthermore, all lengths involved are a fraction of 
the distance $l$ among the primaries, and hence we set
\begin{equation}
\psi \equiv {r \over l}, \; \rho \equiv {\beta \over \alpha}, \; 
\rho_{\alpha} \equiv {l_{\alpha}\over l}, \; \rho_{\beta} \equiv {l_{\beta}\over l}, 
\; \rho_{P} \equiv {l_{P}\over l}.
\label{(3.13)}
\end{equation}
In light of (3.10)-(3.13), we find the following dimensionless form of the nonic resulting 
from Eq. (3.9):
\begin{equation}
\sum_{n=0}^{9}A_{n}\psi^{n}=0,
\label{(3.14)}
\end{equation}
where 
\begin{equation}
A_{0} \equiv =-3 (1+\rho)^{-1}\kappa_{2}(\rho_{P})^{2},
\label{(3.15)}
\end{equation}
\begin{equation}
A_{1} \equiv -2(1+\rho)^{-1}\Bigr[\kappa_{1}\rho_{\alpha}-6 \varepsilon \kappa_{2}(\rho_{P})^{2}\Bigr],
\label{(3.16)}
\end{equation}
\begin{equation}
A_{2} \equiv -(1+\rho)^{-1}\Bigr[1-8 \varepsilon \kappa_{1}\rho_{\alpha}
+18 \kappa_{2} (\rho_{P})^{2}\Bigr],
\label{(3.17)}
\end{equation}
\begin{equation}
A_{3} \equiv 4(1+\rho)^{-1}\Bigr[\varepsilon -3 \kappa_{1}\rho_{\alpha}
+3 \varepsilon \kappa_{2}(\rho_{P})^{2}\Bigr],
\label{(3.18)}
\end{equation}
\begin{equation}
A_{4} \equiv -(1+\rho)^{-1}\Bigr[(6+(1+\varepsilon)\rho)
-2 \kappa_{1}(4 \rho_{\alpha}+\rho_{\beta}\rho)\varepsilon
+3(1+\rho)\kappa_{2}(\rho_{P})^{2}\Bigr],
\label{(3.19)}
\end{equation}
\begin{equation}
A_{5} \equiv (1+\rho)^{-1}\Bigr[(1+4 \varepsilon) +(5+ 2 \varepsilon)\rho
-2 \kappa_{1}(\rho_{\alpha}+\rho_{\beta}\rho)\Bigr],
\label{(3.20)}
\end{equation}
\begin{equation}
A_{6} \equiv -(1+\rho)^{-1}\Bigr[(1+4 \varepsilon)+(10 \varepsilon+1)\rho\Bigr],
\label{(3.21)}
\end{equation}
\begin{equation}
A_{7} \equiv 2(1+\rho)^{-1}(3+5 \rho),
\label{(3.22)}
\end{equation}
\begin{equation}
A_{8} \equiv -(1+\rho)^{-1}(4+5 \rho)\varepsilon ,
\label{(3.23)}
\end{equation}
\begin{equation}
A_{9} \equiv 1.
\label{(3.24)}
\end{equation}
If we take instead the root $s=-(r-\varepsilon l)$ in Eq. (3.7) and insert it into Eq. (3.8), we find,
with analogous procedure, the nonic equation
\begin{equation}
\sum_{n=0}^{9}B_{n}\psi^{n}=0,
\label{(3.25)}
\end{equation}
where
\begin{equation}
B_{k}=A_{k} \; {\rm if} \; k=0,1,2,3,7,8,9,
\label{(3.26)}
\end{equation}
\begin{equation}
B_{4} \equiv (1+\rho)^{-1}\Bigr[(-6+(1-\varepsilon)\rho)
+2 \kappa_{1}\varepsilon (4 \rho_{\alpha}+\rho_{\beta}\rho)
+3(\rho-1)\kappa_{2}(\rho_{P})^{2}\Bigr],
\label{(3.27)}
\end{equation}
\begin{equation}
B_{5} \equiv (1+\rho)^{-1}\Bigr[(1+4 \varepsilon)+(5 -2 \varepsilon)\rho 
-2 \kappa_{1}(\rho_{\alpha}+\rho_{\beta}\rho)\Bigr],
\label{(3.28)}
\end{equation}
\begin{equation}
B_{6} \equiv -(1+\rho)^{-1}\Bigr[(1+4 \varepsilon)+(10 \varepsilon -1)\rho].
\label{(3.29)}
\end{equation}

In Newtonian theory, the collinear Lagrangian points $L_{1},L_{2},L_{3}$ are ruled instead by a
quintic equation, as is clear by setting $k_{1}=k_{2}=k_{3}=0$ in Eq. (3.8)
and multiplying the resulting equation by $(r-\varepsilon l)^{2}r^{2}$. By virtue of the two
choices of sign in Eq. (3.7) one gets, if $s=r -\varepsilon l$, the quintic
\begin{eqnarray}
\sum_{n=0}^{5}C_{n}\psi^{n}&=& \psi^{5}-{(2+ 3 \rho)\over (1+\rho)}\varepsilon \psi^{4}
+{(1+3 \rho)\over (1+\rho)}\psi^{3}
-{[1+(1+ \varepsilon)\rho]\over (1+\rho)}\psi^{2}
\nonumber \\
&+& {2 \varepsilon \over (1+\rho)}\psi -{1 \over (1+\rho)}=0,
\label{(3.30)}
\end{eqnarray}
while $s=-(r- \varepsilon l)$ leads to the quintic
\begin{eqnarray}
\sum_{n=0}^{5}D_{n}\psi^{n}&=& \psi^{5}-{(2+ 3 \rho)\over (1+\rho)}\varepsilon \psi^{4}
+{(1+3 \rho)\over (1+\rho)}\psi^{3}
-{[1-(1- \varepsilon)\rho]\over (1+\rho)}\psi^{2}
\nonumber \\
&+& {2 \varepsilon \over (1+\rho)}\psi -{1 \over (1+\rho)}=0.
\label{(3.31)}
\end{eqnarray}
In this case, the coefficients are related by
\begin{equation}
C_{k}=D_{k} \; {\rm if} \; k=0,1,3,4,5,
\label{(3.32)}
\end{equation}
\begin{equation}
C_{2,-}=-(1+\rho)^{-1}=D_{2,+},
\label{(3.33)}
\end{equation}
\begin{equation}
C_{2,+}=-(1+\rho)^{-1}(1+2 \rho) \not = D_{2,-}=-(1+\rho)^{-1}(1-2 \rho).
\label{(3.34)}
\end{equation}
In light of Eqs. (3.30) and (3.31), we find the values of $r$ (distance from the planetoid
to the primary) in Newtonian theory given by
\begin{equation}
r_{1}=3.2637629578162163 \cdot 10^{8} \; {\rm m},
\label{(3.35)}
\end{equation}
\begin{equation}
r_{2}=3.8167471682615924 \cdot 10^{8} \; {\rm m},
\label{(3.36)}
\end{equation}
\begin{equation}
r_{3}=4.4892055063051933 \cdot 10^{8} \; {\rm m},
\label{(3.37)}
\end{equation}
while, for the corresponding roots of the nonic equations (3.14) and (3.25), we find
\begin{equation}
R_{1}=3.263762957852764 \cdot 10^{8} \; {\rm m},
\label{(3.38)}
\end{equation}
\begin{equation}
R_{2}=3.8167471683504695 \cdot 10^{8} \; {\rm m},
\label{(3.39)}
\end{equation}
\begin{equation}
R_{3}=4.4892055063281494 \cdot 10^{8} \; {\rm m}.
\label{(3.40)}
\end{equation}
By virtue of these values, we find
\begin{equation}
R_{1}-r_{1}=3.6 \; {\rm mm},
\label{(3.41)}
\end{equation}
\begin{equation}
R_{2}-r_{2}=9 \; {\rm mm},
\label{(3.42)}
\end{equation}
\begin{equation}
R_{3}-r_{3}=2.3 \; {\rm mm}.
\label{(3.43)}
\end{equation}
Interestingly, the order of magnitude of quantum corrections to the location of $L_{1},L_{2},L_{3}$ in
the Earth-Moon system coincides with the order of magnitude of quantum corrections to $L_{4},L_{5}$
that we have found in Eq. (2.38). This may not have any practical consequence, since 
$L_{1},L_{2},L_{3}$ are points of unstable equilibrium, but the detailed analysis performed in this
section adds evidence in favour of our evaluation of quantum corrections to all Lagrangian points
in the Earth-Moon system being able to predict effects of order half a centimeter.

The main perturbations of such a scheme may result from the Sun. If one then considers a restricted
four-body problem where the Earth and Moon move in circular orbits around their center of mass, which
in turn moves in a circular orbit about the Sun\footnote{The Sun's effect on the planetoid is much
larger than the Sun's effect on the Moon.}, one finds that $L_{4}$ and $L_{5}$ are no longer points 
of stable equilibrium \cite{Tapley}. However, one can evaluate the impulse required to induce stability
at $L_{4}$, i.e. to force the planetoid to stay precisely at $L_{4}$. Such an impulse turns out to be
$2360$ lb/sec/slug/yr, as shown in Ref. \cite{Tapley}.   

\section{Tiny departure from the equilateral triangle picture: prospects to measure the effect with 
laser ranging}

The quantum gravity effect described in our paper can be studied with the technique of
Satellite/Lunar Laser Ranging (hereafter SLR/LLR) and a laser-ranged test mass equipped with 
Cube Corner Retro-reflectors (CCRs), to be designed ad hoc for this purpose. SLR/LLR is performed by 
the International Laser Ranging Service (ILRS) \cite{Altamimi}, which recently celebrated the 50th
anniversary of the first successful SLR measurement, which occurred at the Goddard Geophysical
and Astronomical Observatory (GGAO) on October 31, 
1964\footnote{See http://ilrs.gsfc.nasa.gov for a description of satellite/lunar laser ranging.
See also http://ilrs.gsfc.nasa.gov/ilrw19/.}. Detecting this tiny departure from classical gravity
is a challenging task, which requires precise positioning in space 
at the Lagrangian points $L_{4}$ and $L_{5}$, in
absolute terms, that is, with respect to an appropriately chosen coordinate reference system. One
potential choice is the International Terrestrial Reference system (ITRS) \cite{Pearlman}, which is
established with several geodesy techniques, including SLR/LLR. The latter provides almost uniquely
the metrological definition of the Earth's center of mass (geocenter) and origin of the ITRS, as 
well as, together with Very Long Baseline Interferometry (VLBI), the absolute scale of length in
space in Earth Orbit. Given its similarity with LLR \cite{Bender}, another option for the coordinate
frame is the Solar System Barycenter (SSB). In fact the distance 
of $L_{4}$ and $L_{5}$ from the ground laser stations 
of the ILRS is very close to their distance to the Laser retroreflector array (LRAs) of deployed by
the Apollo and Lunokhod missions, which over the last 45 years were used for some among the best precision
tests of General Relativity (see Refs. \cite{Williams,Shapiro,Daa,Martini,Marcha,Marchb}). 
The SSB is particularly apt for the purpose, since it is used for General Relativity tests 
carried out with LLR data analysis by means of the orbit software package Planetary Ephemeris
Program (PEP) since the eighties \cite{Shapiro} and until nowadays \cite{Daa,Martini}. PEP 
has been developed by the Harvard-Smithsonian Center for Astrophysics (I. I. Shapiro et al, 
currently maintained by J. F. Chandler). A review of LLR data taking and analysis can be found in
Ref. \cite{Martini}.

A laser ranging test mass (${\rm LR}_{{\rm tm}}$) can be designed with a dedicated effort, by exploiting the 
experience of LLR data taking and analysis described above, and especially by taking advantage
of existing capabilities for detailed pre-launch characterization of any kind of LRAs and/or test
mass for Solar System exploration \cite{Dab,Dac,Currie,Dad}. Some of the Key Performance 
Indicators (KPIs) that must be taken into account to design an appropriate ${\rm LR}_{{\rm tm}}$ for the
signature of new physics described in this paper are as follows.
\vskip 0.3cm
\noindent
(i) Adequate laser return signal (lidar optical cross section) from the Lagrangian 
points $L_{4},L_{5}$.
\vskip 0.3cm
\noindent
(ii) Acceptable rejection of the unavoidable nongravitational perturbations (NGPs) 
at $L_{4},L_{5}$ which {\it any} chosen test mass and/or 
test spacecraft will experience, whose complexity scales with
the complexity of the structure of the test mass and/or test spacecraft itself.
\vskip 0.3cm
\noindent
(iii) Optimization/minimization of the value of the surface-to-mass ratio, S/M. This is a critical
KPI, since all NGPs related to the sun radiation pressure and thermal effect, are proportional
to S/M (see for example Ref. \cite{Vok}). Compared to other test spacecrafts and/or test masses an
${\rm LR}_{{\rm tm}}$ has the advantage of the simplicity of 
geometrical shape (for example, spherical) and mechanical
structure. To date, Apollo/Lunokhod are demonstrating a lifetime of at least 45 years.
\vskip 0.3cm
\noindent
(iv) Time-durability of the test mass to prolonged measurements. Since ${\rm LR}_{{\rm tm}}$ are passive and
maintenance free, this KPI favors ${\rm LR}_{{\rm tm}}$ over 
other types of any active test masses and/or spacecrafts.

The above KPIs can be characterized at the dedicated laboratory described by Refs. \cite{Dab,Dac}
(see also http://www.lnf.infn.it/esperimenti/etrusco/). From the experimental point of view of laser ranging
investigations, arguments reported in this section for $L_{4}$ and $L_{5}$ apply identically to $L_{3}$. 
They do not apply to $L_{2}$ since such a position is not visible from ILRS stations. The distance of 
$L_{1}$ from Earth is shorter than for $L_{3},L_{4}$ and $L_{5}$, which would make the laser return signal
from an $LR_{\rm tm}$ in $L_{1}$ higher than from $L_{3},L_{4},L_{5}$ (by a purely geometric factor equal
to the fourth power of the ratio of the distances of $L_{3}$ and $L_{1}$ from any given ILRS station;
see for example Ref. \cite{Dab}). Given the relative proximity of $L_{1}$ to the Moon, gravitational effects
on an $LR_{\rm tm}$ in $L_{1}$ related to the nonpointlike structure of the Moon (felt in $L_{1}$) should be
evaluated to determine their influence, if any, on the conclusions of the previous section. This influence is
expected to be negligible for an $LR_{\rm tm}$ in $L_{3},L_{4}$ and $L_{5}$, since they are much more 
distant from the Moon than $L_{1}$.  

\section{Displaced periodic orbits for a solar sail in the Earth-Moon system}

Displaced periodic orbits describe the dynamics of the planetoid,
e.g., a solar sail, in the neighborhood of the libration points, 
which have been studied in detail in the quantum-corrected case \cite{BE14a} and in Newtonian theory
\cite{Simo08}. The appropriate tool of classical mechanics are the variational equations, for which we
refer the reader to Refs. \cite{P1892,Pars65,BE14b}. In the simplest possible terms, the components
$x,y,z$ of the position vector of the sail (see Fig. 2) 
at each libration point change by the infinitesimal
amount $\xi,\eta,\zeta$ respectively and, by retaining only first-order terms in 
$\xi,\eta,\zeta$ in the equations of motion, one finds the following linear variational equations of
motion for the libration points $L_{4},L_{5}$ describing stable equilibrium \cite{Simo08}
\begin{equation}
\ddot \xi -2 \dot \eta=U_{xx}^{0}\xi + U_{xy}^{0}\eta+a_{\xi},
\label{(5.1)}
\end{equation}
\begin{equation}
\ddot \eta +2 \dot \xi=U_{xy}^{0}\xi + U_{yy}^{0}\eta+a_{\eta},
\label{(5.2)}
\end{equation}
\begin{equation}
\ddot \zeta =U_{zz}^{0}\zeta +a_{\zeta},
\label{(5.3)}
\end{equation}
where the auxiliary variables $a_{\xi},a_{\eta},a_{\zeta}$ describe the solar sail acceleration, and 
$U_{xx}^{0},U_{yy}^{0},U_{zz}^{0},U_{xy}^{0}$ are the partial derivatives of the gravitational potential (1.7) 
evaluated at $L_{4}$ or $L_{5}$. Note that, following Ref. \cite{Simo08}, we are here using units where
the sum of the masses of the primaries is set to $1$, as well as their distance and the Newton constant. 

\begin{figure}
\includegraphics[scale=0.70]{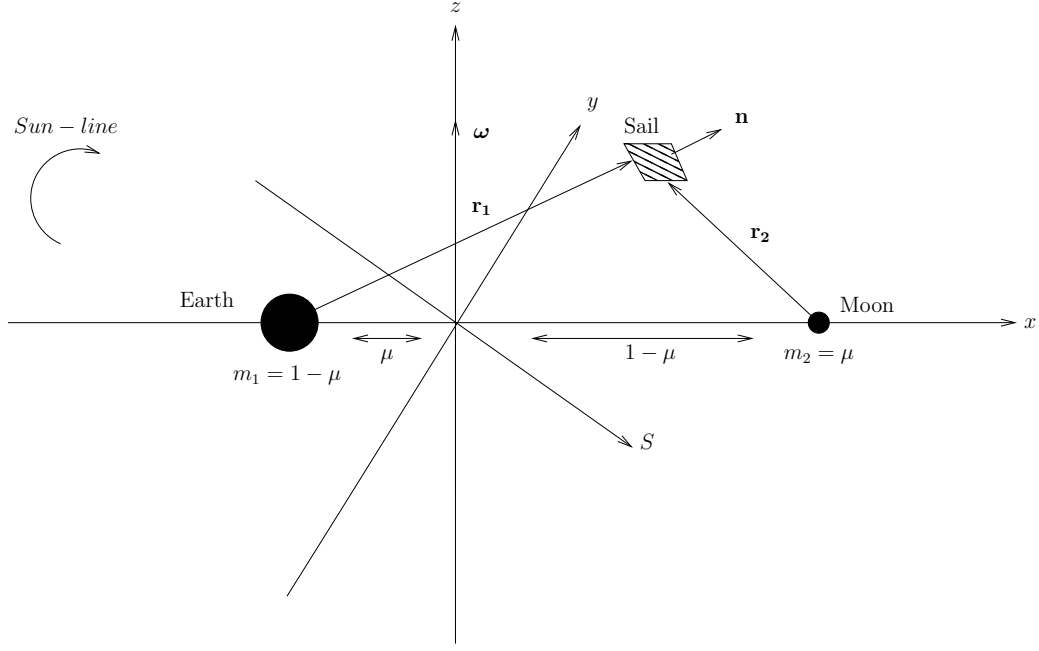}
\caption{Schematic geometry of the Earth-Moon restricted three-body problem when the planetoid is
a solar sail.}
\end{figure}

Following Ref. \cite{Simo08}, assume now that a solution of the linearized equations of motion (5.1)-(5.3)
is periodic of the form
\begin{equation}
\xi(t)=A_{\xi}\cos(\omega_{\star}t)+B_{\xi}\sin(\omega_{\star}t),
\label{(5.4)}
\end{equation}
\begin{equation}
\eta(t)=A_{\eta}\cos(\omega_{\star}t)+B_{\eta}\sin(\omega_{\star}t),
\label{(5.5)}
\end{equation}
where $A_{\xi},A_{\eta},B_{\xi}$ and $B_{\eta}$ are parameters to be determined, and
$\omega_{\star}=0.923$ is the angular rate of the Sun line in the corotating frame in a 
dimensionless synodic coordinate system \cite{Simo08}. By substituting Eqs. (5.4)-(5.5) in the
differential equations (5.1)-(5.3), we obtain the following linear system in 
$A_{\xi},A_{\eta},B_{\xi}$ and $B_{\eta}$ \cite{Simo08}:
\begin{equation}
-(\omega_{\star}^{2}+U_{xx}^{0})B_{\xi}+2 \omega_{\star}A_{\eta}-U_{xy}^{0}B_{\eta}=0,
\label{(5.6)}
\end{equation}
\begin{equation}
-U_{xy}^{0}A_{\xi}+2 \omega_{\star}B_{\xi}-(\omega_{\star}^{2}+U_{yy}^{0})A_{\eta}=0,
\label{(5.7)}
\end{equation}
\begin{equation}
-(\omega_{\star}^{2}+U_{xx}^{0})A_{\xi}-U_{xy}^{0}A_{\eta}-2 \omega_{\star}B_{\eta}=a_{0}\cos^{3}\varphi,
\label{(5.8)}
\end{equation}
\begin{equation}
-2 \omega_{\star}A_{\xi}-U_{xy}^{0}B_{\xi}-(\omega_{\star}^{2}+U_{yy}^{0})B_{\eta}
=-a_{0}\cos^{3}\varphi.
\label{(5.9)}
\end{equation}
This linear system can be solved to find the coefficients $A_{\xi},B_{\xi},A_{\eta},B_{\eta}$, here
arranged in the four rows of a column vector ${\bf P}$, while ${\bf b}$ is the column vector whose four
rows are the right-hand sides of (5.6), (5.7), (5.8) and (5.9), respectively. Let $A$ be the 
$4 \times 4$ matrix
\begin{equation}
A= \left(\begin{matrix} A_{1} & B_{1} \cr C_{1} & D_{1}
\end{matrix}\right),
\label{(5.10)}
\end{equation}
where the $2 \times 2$ submatrices of $A$ are \cite{Simo08}
\begin{equation}
A_{1}=\left(\begin{matrix} 
0 & -\omega_{\star}^{2}-U_{xx}^{0} \cr
-U_{xy}^{0} & 2 \omega_{\star} 
\end{matrix}\right),
\label{(5.11)}
\end{equation}
\begin{equation}
B_{1}=\left(\begin{matrix}
2 \omega_{\star} & -U_{xy}^{0} \cr
-\omega_{\star}^{2}-U_{yy}^{0} & 0
\end{matrix}\right),
\label{(5.12)}
\end{equation}
\begin{equation}
C_{1}=\left(\begin{matrix}
-\omega_{\star}^{2}-U_{xx}^{0} & 0 \cr
-2 \omega_{\star} & -U_{xy}^{0} 
\end{matrix}\right),
\label{(5.13)}
\end{equation}
\begin{equation}
D_{1}=\left(\begin{matrix}
-U_{xy}^{0} & -2 \omega_{\star} \cr
0 & -\omega_{\star}^{2}-U_{yy}^{0}
\end{matrix}\right).
\label{(5.14)}
\end{equation}
With this matrix notation, the solution of our linear system (5.6)-(5.9) reads as \cite{Simo08}
\begin{equation}
P^{i}=(A^{-1})_{\; j}^{i} \; b^{j} \; \forall i=1,2,3,4.
\label{(5.15)}
\end{equation}
The coefficients $A_{\xi},A_{\eta},B_{\xi}$ and $B_{\eta}$ are amplitudes that characterize the
displaced periodic orbit.

Last, the out-of-plane motion (Eq. (4.3)) is decoupled from the in-plane motion, hence the solution
of Eq. (5.3) is given by \cite{Simo08}
\begin{eqnarray}
\zeta(t)&=& \theta(t) a_{0}\cos^{2}\varphi (\sin \varphi)|U_{zz}^{0}|^{-1}
\nonumber \\
&+& \cos(\omega_{\zeta}t)\Bigr[\zeta(t=0)-a_{0}\cos^{2}\varphi (\sin \varphi)
|U_{zz}^{0}|^{-1}\Bigr],
\label{(5.16)}
\end{eqnarray}
where $\theta$ is a step function 
\begin{equation}
\theta(t)=1 \; {\rm if} \; t>0, \; \theta(t)=0 \; {\rm if} \; t<0.
\label{(5.17)}
\end{equation} 
Thus, the required sail acceleration for a fixed distance can be given by \cite{Simo08}
\begin{equation}
a_{0}={\zeta(t=0)|U_{zz}^{0}| \over \cos^{2}\varphi (\sin \varphi)}.
\label{(5.18)}
\end{equation}

In Newtonian theory, the findings for displaced periodic orbits are well summarized in Figs.
3-5 (cf. Ref. \cite{Simo08}).

\begin{figure}
\includegraphics[scale=0.70]{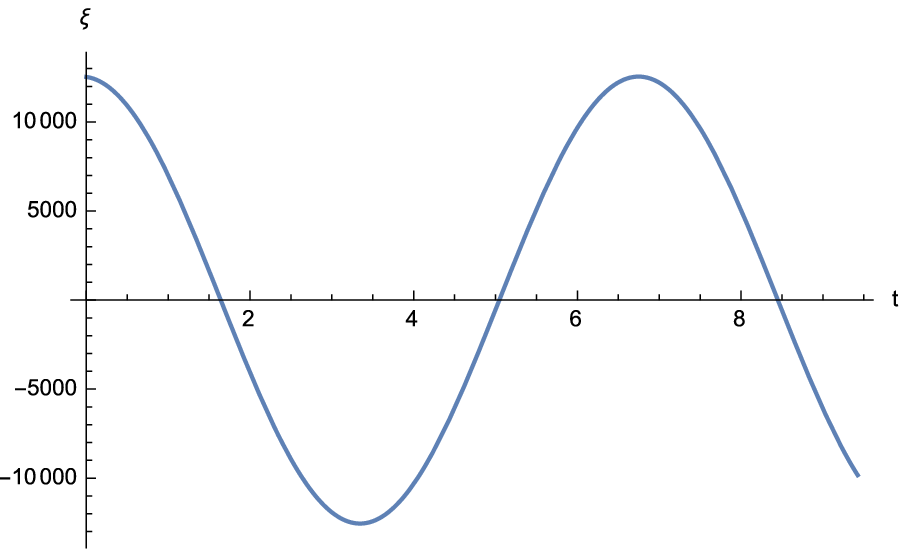}
\caption{Time evolution of $\xi$ for $L_{4}$ in the Newtonian case.}
\end{figure}

\begin{figure}
\includegraphics[scale=0.70]{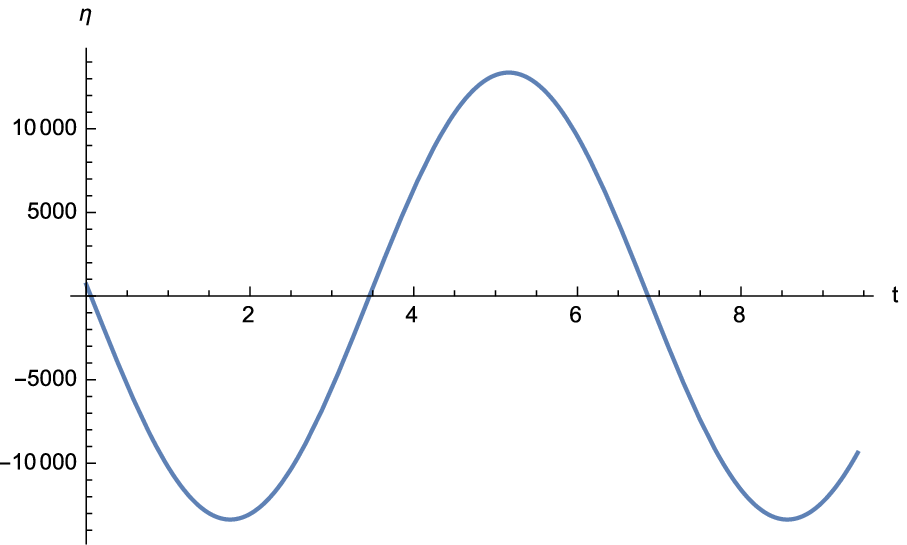}
\caption{Time evolution of $\eta$ for $L_{4}$ in the Newtonian case.}
\end{figure}

\begin{figure}
\includegraphics[scale=0.70]{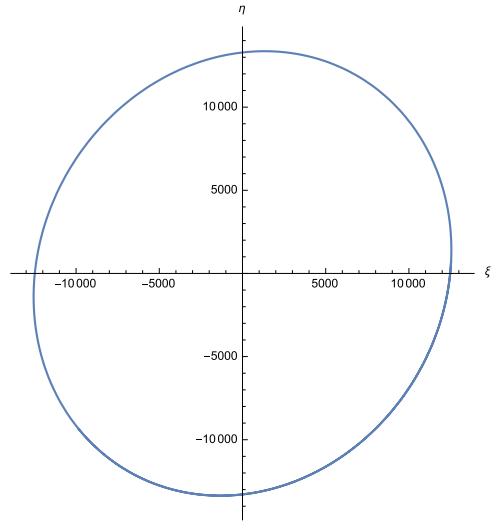}
\caption{Periodic orbits at linear order around the Lagrangian point $L_{4}$ in Newtonian theory.}
\end{figure}

On considering the quantum corrections evaluated in detail in Sec. II, and setting furthermore
the angle $\varphi={\pi \over 4}$, while $\zeta(t=0)=2$, $\omega_{\star}=0.923$, $a_{0}=10^{-4}$,
we arrive at the plots displayed in Figs. 6-8. The starting value of $\zeta$ has been taken to be $100$ km,
increased gradually to reach $2500$ km.

\begin{figure}
\includegraphics[scale=0.70]{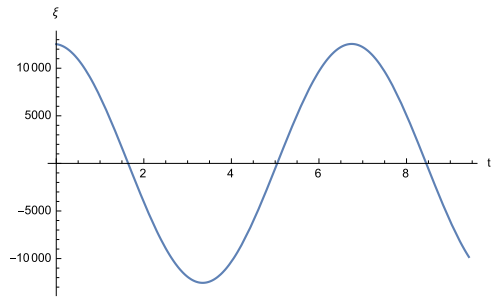}
\caption{Time evolution of $\xi$ for $L_{4}$ in the quantum-corrected model.}
\end{figure}

\begin{figure}
\includegraphics[scale=0.70]{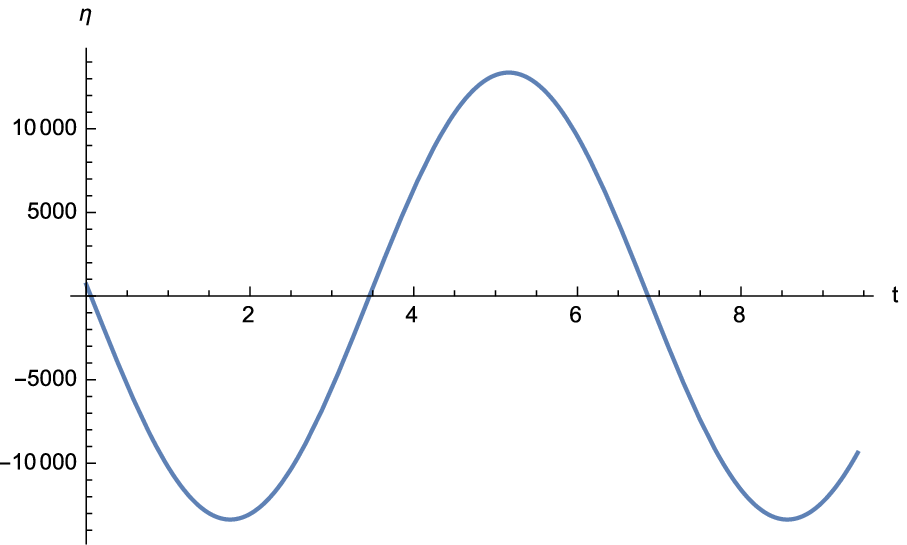}
\caption{Time evolution of $\eta$ for $L_{4}$ in the quantum-corrected model.}
\end{figure}

\begin{figure}
\includegraphics[scale=0.70]{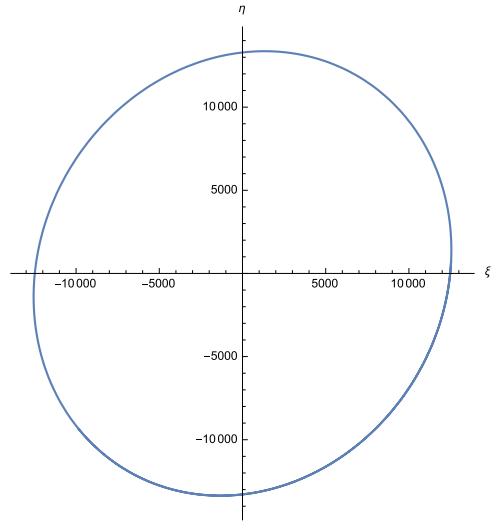}
\caption{Periodic orbits at linear order around the Lagrangian point $L_{4}$. We have used the 
quantum corrected coordinates obtained at the end of Sec. II. The periodic orbit is elliptic as
in the Newtonian case displayed in Fig. 5.}
\end{figure}

Our calculation is of interest because it shows that even our quantum corrected potential allows
for periodic solutions in the neighborhood of uniform circular motion. 
The precise characterization of regions of stability and instability \cite{LeviCivita} 
of such displaced periodic orbits is a fascinating problem for the years to come.

\section{Concluding remarks and open problems}

We find it appropriate to begin our concluding remarks by stressing two conceptual aspects, which
are as follows. 
\vskip 0.3cm
\noindent
(i) In the course of an orbit of a celestial body around another celestial body, their mutual
separation may change by a nonnegligible amount. Thus, it would be misleading to look for an
observational test of one-loop long-distance quantum corrections to the Newtonian potential by investigating
the orbits, because we do not have a formula for $V(r)$ which is equally good at all points.
By contrast, the evaluation of stable equilibrium points (to first order in perturbations)
provides a definite prediction, i.e., the coordinates of such a point, which can be 
hopefully measured with the techniques outlined in Sec. III. In other words, coordinates of
Lagrangian libration points $L_{4}$ and $L_{5}$ and displaced periodic orbits around 
unperturbed circular motion provide a valuable
test of effective field theories of quantum gravity, whereas the orbits of celestial bodies are best
studied within the framework of relativistic celestial mechanics. 
\vskip 0.3cm
\noindent
(ii) At the risk of repeating ourselves, the technique of Refs. \cite{D94,D94b,D94c,D03} provides 
corrections to the Newtonian potential, and hence the unperturbed dynamics is the Newtonian
celestial mechanics of the Earth-Moon-satellite system, which may provide a good example of
circular restricted three-body problem. The quantum corrected potential becomes (1.2), where
$k_{1}$ appears, on dimensional ground, purely classical, but includes a numerical coefficient,
$\kappa_{1}$, which depends on the value taken by the coefficient $\kappa_{2}$ that
multiplies $l_{P}^{2}$ in $k_{2}$:
$$
\kappa_{1}=\kappa_{1}(\kappa_{2}), \;
k_{2}=\kappa_{2}l_{P}^{2}.
$$
Thus, we do not compute corrections to relativistic celestial mechanics
(cf. Ref. \cite{Huang14}), but, on the other hand, we need the advanced tools of 
relativistic celestial mechanics to test the tiny effect predicted in Eq. (2.38).
We should mention at this stage the important work in Ref. \cite{Yamada12}, where the authors
obtain a triangular solution to the general relativistic three-body problem for general
masses, and find that the post-Newtonian configuration for three finite masses is not 
always equilateral. When their technique is applied to the Earth-Moon system, we find,
unlike our Eq. (2.38), a correction to the $x$-coordinate of $L_{4}$ of order $2.73$ mm, and a
correction to the $y$ coordinate of $L_{4}$ of order $-0.53$ mm. The former agrees with our
orders of magnitude, while the latter, being less than a millimeter, is very hardly detectable.
 
Our original contribution is, first, the detailed calculation of the roots of the quintic equation
in Sec. II (which is an original application of techniques previously developed by mathematicians),
second, the derivation and solution of the nonic equation in Sec. III for quantum corrections to
collinear Lagrangian points and, third, the application of the roots 
in Sec. II to the evaluation of displaced periodic
orbits of solar sails in the Earth-Moon system, when the one-loop long distance quantum corrections
to the Newtonian potentials are taken into account \cite{D03,BE14a,BE14b}. The use of dimensionless
variables for the quintic equation (2.6), and the exploitment of exact formulas for its roots
were of crucial importance to double-check the numerical predictions of Refs. \cite{BE14a,BE14b}. 
Interestingly, we have found that a refined analysis like ours confirms the orders of magnitude
obtained in Ref. \cite{BE14b}, whereas the sign of $|y_{Q}|-|y_{C}|$ gets reversed with respect
to Ref. \cite{BE14b}, and its expected theoretical value turns out to be smaller by $20$ per cent. 
Furthermore, displaced periodic orbits have been evaluated in Sec. IV with the quantum corrected
coordinates displayed in Eq. (2.36), when the condition for the existence of displaced orbits is
affected by terms resulting from a solar sail model. We have found that, even when the quantum 
corrected potential (1.2) is adopted, the displaced periodic orbits are of elliptical shape
(see Fig. 8) as in the Newtonian theory. The solar-sail model is an interesting possibility
considered over the last few dacades, but is not necessarily better than alternative models of 
planetoid. For example, the large structure and optical nature of solar sails can create a
considerable challenge. If the structure and mass distribution of the sail is complicated, one has
to resort to suitable approximations. Furthermore, the characterization of regions of stability
or instability \cite{LeviCivita} of displaced periodic orbits of solar sails is a theoretical
problem whose solution might have far reaching consequences for designing space missions.

Last, but not least, the laser ranging techniques outlined in Sec. IV appear as a promising tool
for testing the predictions at the end of Sec. II. The years to come will hopefully tell us whether
a laser ranging test mass can be designed, upon consideration of the four key performance indicators
listed at the end of Sec. IV. At that stage, the task will remain to actually send a satellite at 
$L_{4}$ and keep it there despite the perturbations caused by the Sun \cite{Tapley}, 
which has never been accomplished to the best of our knowledge. The resulting low-energy
test of quantum gravity in the solar system would reward the considerable effort necessary
to achieve this. 

\acknowledgments 
E. B. and G. E.  are indebted to John Donoghue for enlightening correspondence, and to Massimo Cerdonio
and Alberto Vecchiato for conversations. 
G. E. is grateful to the Dipartimento di Fisica of
Federico II University, Naples, for hospitality and support. 

\begin{appendix}

\section{Bring-Jerrard form of quintic equations}

Let us start from the general quintic equation
\begin{equation}
X^{5}+a_{4}X^{4}+a_{3}X^{3}+a_{2}X^{2}+a_{1}X+a_{0}=0.
\label{(A1)}
\end{equation}
Denote the roots of Eq. (A1) by $X_{i},i=1,...,5$, and let
\begin{equation}
S_{n}=S_{n}(X_{k}) \equiv \sum_{k=1}^{5}(X_{k})^{n}
\label{(A2)}
\end{equation}
be the sum of the $n$th powers of such roots. By virtue of the Newton power-sum formula, a general
representation of $S_{n}$ is
\begin{equation}
S_{n}=-n a_{5-n}-\sum_{j=1}^{n-1}S_{n-j}a_{5-j},
\label{(A3)}
\end{equation}
with the understanding that $a_{j}=0$ for $j<0$. For the lowest values of $n$, Eq. (A3) yields
$S_{1}(X_{k})=-a_{4}, \; S_{2}(X_{k})=(a_{4})^{2}-2a_{3}, \; 
S_{3}(X_{k})=-(a_{4})^{3}+3a_{3}a_{4}-3a_{2}$, 
\begin{equation}
S_{4}(X_{k})=(a_{4})^{4}-4a_{3}(a_{4})^{2}+4 a_{2}a_{4}+2 (a_{3})^{2}-4 a_{1},
\label{(A4)}
\end{equation}
\begin{equation}
S_{5}(X_{k})=-(a_{4})^{5}+5\Bigr[a_{3}(a_{4})^{3}-a_{2}(a_{4})^{2}-(a_{3})^{2}a_{4}+a_{1}a_{4}-a_{0}
+a_{2}a_{3}\Bigr].
\label{(A5)}
\end{equation}
A systematic way to proceed involves two steps, i.e., first a quadratic Tschirnhaus transformation \cite{T}
\begin{equation}
Y_{k}=(X_{k})^{2}+\mu X_{k}+\nu
\label{(A6)}
\end{equation}
between the roots $X_{k}$ of Eq. (A1) and the roots $Y_{k}$ of the principal quintic 
\begin{equation}
X^{5}+c_{2}X^{2}+c_{1}X+c_{0}=0,
\label{(A7)}
\end{equation}
supplemented \cite{Adamchik} by the evaluation of $S_{1}(Y_{k}),...,S_{5}(Y_{k})$ to obtain through
radicals $\mu,\nu,c_{0},c_{1},c_{2}$, and eventually a quartic Tschirnhaus transformation \cite{T}
\begin{equation}
Z_{k}=(Y_{k})^{4}+u_{1}(Y_{k})^{3}+u_{2}(Y_{k})^{2}+u_{3}Y_{k}+u_{4},
\label{(A8)}
\end{equation}
between the roots $Y_{k}$ of Eq. (A7) and the roots $Z_{k}$ of the Bring-Jerrard form (2.10) of the 
quintic. This procedure is conceptually clear although rather lengthy (see Appendix B), and the joint
effect of inverting (A8) and then (A6) to find $X_{k}=X_{k}(Y_{j}(Z_{l}))$ leads to $20$ candidate
roots \cite{Adamchik}, which is not very helpful if one is interested in the numerical values of
such roots, as indeed we are.
One might instead stick to our Eq. (2.6) and then try to exploit the Birkeland theorem
\cite{Birkeland1920}, according to which the roots of any quintic can be re-expressed through generalized
hypergeometric functions like the ones defined in our Sec. II. However, when there are three (or more)
nonvanishing coefficients as in Eq. (2.6), the Birkeland theorem leads to too many 
(for numerical purposes) hypergeometric functions in the general expansion of roots. Furthermore,
the set of linear partial differential equations \cite{Sturmfels} obeyed by the roots when viewed as
functions of all coefficients does not lead easily to their explicit form. 

At this stage, having appreciated the need for a trinomial, possibly Bring-Jerrard form of the
quintic, and rather than feeling in despair, we point out that, 
since in our original quintic (2.6) two coefficients vanish, i.e. $a_{2}=a_{1}=0$, it is more
convenient to use what is normally ruled out in the generic case \cite{Adamchik}, i.e., a cubic
Tschirnhaus transformation between the roots $X_{k}$ of Eq. (A1) and the roots $Y_{k}$ of Eq. (2.10):
\begin{equation}
Y_{k}=(X_{k})^{3}+\lambda_{1}(X_{k})^{2}+\lambda_{2}(X_{k})+\lambda_{3}.
\label{(A9)}
\end{equation}
By virtue of Eqs. (2.10) and (A2)-(A5), we find 
\begin{equation}
S_{1}(Y_{k})=S_{2}(Y_{k})=S_{3}(Y_{k})=0, 
\label{(A10)}
\end{equation}
\begin{equation}
S_{4}(Y_{k})=-4 d_{1}, \;
S_{5}(Y_{k})=-5 d_{0}.
\label{(A11)}
\end{equation}
On assuming the cubic relation (A9), Eqs. (A10) become a nonlinear algebraic system leading to the
numerical evaluation of $\lambda_{1},\lambda_{2},\lambda_{3}$. More precisely, from 
$S_{1}(Y_{k})=0$ we find
\begin{equation}
5\lambda_{3}+\lambda_{2}S_{1}(X_{k})+\lambda_{1}S_{2}(X_{k})+S_{3}(X_{k})=0,
\label{(A12)}
\end{equation}
while from $S_{2}(Y_{k})=0$ we obtain
\begin{eqnarray}
\; & \; &
5(\lambda_{3})^{2}+2 \lambda_{2}\lambda_{3}S_{1}(X_{k})
+[(\lambda_{2})^{2}+2 \lambda_{1}\lambda_{3}]S_{2}(X_{k})
+2(\lambda_{1}\lambda_{2}+\lambda_{3})S_{3}(X_{k}) 
\nonumber \\
&+& [(\lambda_{1})^{2}+2 \lambda_{2}]S_{4}(X_{k})
+2 \lambda_{1}S_{5}(X_{k})+S_{6}(X_{k})=0.
\label{(A13)}
\end{eqnarray}
Last, from the vanishing of $S_{3}(Y_{k})$ we get
\begin{eqnarray}
\; & \; &
5 (\lambda_{3})^{3}+3 \lambda_{2}(\lambda_{3})^{2}S_{1}(X_{k})
+3 (\lambda_{2})^{2}\lambda_{3}S_{2}(X_{k})+(\lambda_{2})^{3}S_{3}(X_{k})
\nonumber \\
&+& 3 (\lambda_{1})^{2}\lambda_{3}S_{4}(X_{k})
+[3(\lambda_{1})^{2}\lambda_{2}+6 \lambda_{1}\lambda_{3}]S_{5}(X_{k})
+[(\lambda_{1})^{3}+3 \lambda_{3}+6 \lambda_{1}\lambda_{2}]S_{6}(X_{k})
\nonumber \\
&+& 3[(\lambda_{1})^{2}+\lambda_{2}]S_{7}(X_{k})
+3 \lambda_{1}S_{8}(X_{k})+S_{9}(X_{k})=0.
\label{(A14)}
\end{eqnarray}
The system (A12)-(A14) cannot be solved by radicals because, if one expresses for example
$\lambda_{1}$ as a linear function of $\lambda_{2}$ and $\lambda_{3}$ from Eq. (A12), and one
solves the resulting quadratic equation for $\lambda_{2}=\lambda_{2}(\lambda_{3})$ or 
$\lambda_{3}=\lambda_{3}(\lambda_{2})$ from Eq. (A13), 
one discovers that Eq. (A14) is not a polynomial
in $\lambda_{3}$ (respectively, $\lambda_{2}$). Nevertheless, {\it for numerical purposes},
the system (A12)-(A14) can be solved and has been solved from us in the Earth-Moon system.
Last, from Eq. (A11) we find the coefficients $d_{1}$ and $d_{0}$ in the Bring-Jerrard form of
the quintic, according to the formulas
\begin{equation}
d_{1}=-{1 \over 4}S_{4}(Y_{k})=\sum_{i=0}^{12}b_{1i}S_{i}(X_{k}),
\label{(A15)}
\end{equation}
\begin{equation}
d_{0}=-{1 \over 5}S_{5}(Y_{k})=\sum_{i=0}^{15}b_{0i}S_{i}(X_{k}),
\label{(A16)}
\end{equation}
We have evaluated all $b_{1i}$ and $b_{0i}$ coefficients by applying patiently the Tschirnhaus 
transformation (A9) and the definition (A2). We find therefore six triplets of possible values 
for $\lambda_{1},\lambda_{2},\lambda_{3}$ (see Tables I and II), 
which lead always to the same values of 
$d_{1}$ and $d_{0}$ (this is a crucial consistency check), i.e.
\begin{equation}
d_{1}(w)=2.78 \cdot 10^{-176}+{\rm i} 6.91 \cdot 10^{-186}, \;
d_{1}(u)=2.78 \cdot 10^{-176}-{\rm i}8.51 \cdot 10^{-188},
\label{(A17)}
\end{equation}
\begin{equation}
d_{0}(w)=1.66 \cdot 10^{-220}-{\rm i} 5.17 \cdot 10^{-230}, \;
d_{0}(u)=-1.66 \cdot 10^{-220}+{\rm i} 6.36 \cdot 10^{-232},
\label{(A18)}
\end{equation}
where $w$ and $u$ are the variables defined in Eqs. (2.5) and (2.32), respectively.
Eventually, the roots $X_{k}$ of our quintic (2.6) have been obtained by solving Eq. (A9) for
$X_{k}=X_{k}(Y_{k})$, with the help of the solution algorithm for the cubic equation.
This means that we first re-express (A9) in the form
\begin{equation}
h(X_{k}) \equiv (X_{k})^{3}+\kappa_{2}(X_{k})^{2}+\kappa_{1}X_{k}+\kappa_{0}=0,
\label{(A19)}
\end{equation}
where $\kappa_{2} \equiv \lambda_{1}, \kappa_{1} \equiv \lambda_{2}, 
\kappa_{0} \equiv \lambda_{3}-Y_{k}$. We then define the new variable
\begin{equation}
V_{k} \equiv X_{k}+{\kappa_{2} \over 3}=X_{k}+{\lambda_{1} \over 3},
\label{(A20)}
\end{equation}
in terms of which Eq. (A19) is mapped into its canonical form
\begin{equation}
(V_{k})^{3}+pV_{k}+q=0, \; p \equiv h' \left(-{\kappa_{2}\over 3}\right), \;
q \equiv h \left(-{\kappa_{2}\over 3}\right).
\label{(A21)}
\end{equation}
As shown in Ref. \cite{Birkeland1924}, if the discriminant
\begin{equation}
\delta \equiv -{27 \over 4}{q^{2}\over p^{3}}
\label{(A22)}
\end{equation}
is such that $|\delta|<1$, or if $\delta=1$, the three roots of Eq. (A21) can be expressed through the
Gauss hypergeometric function in the form
\begin{equation}
(V_{k})_{i}=\sqrt{-p}\left[(-1)^{3i}F \left(-{1 \over 6},{1 \over 6}, {1 \over 2}; \delta \right)
+{1 \over 3}\sqrt{\delta \over 3}F \left({1 \over 3},{2 \over 3},{3 \over 2}; \delta \right)\right]
(i=1,2), 
\label{(A23)}
\end{equation}
\begin{equation}
(V_{k})_{3}=-{2 \over 3}\sqrt{-{p \delta \over 3}} \; 
F \left({1 \over 3},{2 \over 3},{3 \over 2}; \delta \right).
\label{(A24)}
\end{equation}
As is clear from (A20), (A23), (A24), our method yields eventually $15$ candidate roots, and by
insertion into the original quintic (2.6) we have found the $5$ effective roots at the
end of Sec. II.

\begin{table}[!h]
\caption{The six triplets of values of  $\lambda_3$, $\lambda_2$ and $\lambda_1$ 
for the $1/r$-equation (2.1).}
\begin{tabular}{|c|c|c|c|}
\hline
$n$th triplet & $\lambda_3$  & $\lambda_2$  &   $\lambda_1$   \\
\hline
n=1 &  $-4.98 \times 10^{45}-3.10 \times 10^{-55} \; \mathrm{i} $ &  $ 0.26 +  1.59\times10^{-11}\;  
\mathrm{i}  $ &  $4.21 \times 10^{32} + 3.78 \times 10^{-44} \; \mathrm{i}$ \\
\hline
n=2 &  $ -4.98 \times 10^{-45} + 3.10 \times 10^{-55} \; \mathrm{i} $ &  $ 0.26 -  1.59 \times 10^{-11} \; 
\mathrm{i} $ & $4.21\times10^{32} - 3.78 \times 10^{-44} \; \mathrm{i} $ \\
\hline
n=3 & $2.49 \times 10^{-45} + 4.32\times 10^{-45} \; \mathrm{i} $ &  $ 0.26 + 2.57 \times 10^{-11} \; 
\mathrm{i} $&   $ 4.21\times 10^{32} +  6.11 \times 10^{-44} \; \mathrm{i} $\\
\hline
n=4 &  $ 2.49\times 10^{-45} - 4.32\times 10^{-45} \; \mathrm{i} $ & $ 0.26 - 2.57 \times 10^{-11} \; 
\mathrm{i} $ &  $ 4.21\times 10^{32} - 6.11 \times 10^{-44} \; \mathrm{i}$ \\
\hline
n=5 & $2.49 \times 10^{-45} + 4.32 \times 10^{-45} \; \mathrm{i}$ &  $0.26 +  9.82 \times 10^{-12} \; 
\mathrm{i}$ & $ 4.21 \times 10^{32} +  2.34 \times 10^{-44} \; \mathrm{i} $ \\
\hline
n=6 &  $ 2.49 \times 10^{-45} - 4.32 \times 10^{-45} \; \mathrm{i}$&  $0.26 - 9.82 \times 10^{-12} \; 
\mathrm{i}$ &  $4.21\times 10^{32} - 2.34\times 10^{-44} \; \mathrm{i}$ \\
\hline
\end{tabular}
\end{table} 

\begin{table}[!h]
\caption{The six triplets of values of  $\lambda_3$, $\lambda_2$ and $\lambda_1$ 
for the $1/s$-equation (2.2).}
\begin{tabular}{|c|c|c|c|}
\hline
$n$th triplet & $\lambda_3$  & $\lambda_2$  &   $\lambda_1$   \\
\hline
n=1 &   $-4.98\times 10^{-45} - 3.82\times 10^{-57} \; \mathrm{i}$& $ 0.26 + 1.96\times 10^{-13} \; 
\mathrm{i}$&  $ 5.17\times 10^{30} + 3.78\times 10^{-44} \; \mathrm{i}$ \\
\hline
n=2 & $-4.98\times 10^{-45} + 3.82\times 10^{-57}  \; \mathrm{i} $ &  $ 0.26 -  1.96\times 10{-13}  \; 
\mathrm{i} $ & $ 5.17\times 10^{30} - 3.78\times 10^{-44}  \; \mathrm{i} $ \\
\hline
n=3 & $2.49\times 10^{-45} + 4.32\times 10^{-45} \; \mathrm{i} $ & $0.26 + 3.16\times 10^{-13} \; 
\mathrm{i} $ & $5.17\times 10^{30} + 6.11\times10^{-44} \; \mathrm{i} $\\
\hline
n=4 & $2.49\times 10^{-45} - 4.32\times 10^{-45} \; \mathrm{i}$ & $ 0.26 - 3.16\times 10^{-13} \; 
\mathrm{i} $ & $ 5.17 \times 10^{30} - 6.11\times 10^{-44} \; \mathrm{i}$ \\
\hline
n=5 & $ 2.49\times 10^{-45} + 4.32\times 10^{-45} \; \mathrm{i} $ &  $ 0.26 + 1.21\times 10^{-13} \; 
\mathrm{i} $ & $5.17\times 10^{30} + 2.34\times10^{-44} \; \mathrm{i} $ \\
\hline
n=6 & $ 2.49\times 10^{-45} - 4.32\times 10^{-45} \; \mathrm{i} $ & $0.26 - 1.21\times 10^{-13} \; 
\mathrm{i} $ & $5.17\times 10^{30} - 2.34\times 10^{-44} \; \mathrm{i}  $ \\
\hline
\end{tabular}
\end{table} 

Yet another valuable solution algorithm is available, i.e. the method in Ref. \cite{King}
which expresses the roots of the quintic (A1) through two infinite series, i.e., the Jacobi nome
and the theta series, for which fast convergence is obtained, but the need to evaluate the roots
with a large number of decimal digits makes it problematic, as far as we can see, to deal with
such series. Further valuable work on the quintic can be found in Ref. \cite{Drociuk}.

\section{Alternative route to the quintic}

We here find it useful to give details about the main alternative to the procedure used in
Appendix A. For that purpose, as we said, one assumes that the roots $X_{k}$ of Eq. (A1) 
are related to the roots $Y_{k}$ of the principal quintic (A7) by the quadratic transformation
(A6). The power sums for the principal quintic form are indeed 
\begin{equation}
S_{1}(Y_{k})=S_{2}(Y_{k})=0, \;
S_{3}(Y_{k})=-3c_{2}, \; S_{4}(Y_{k})=-4 c_{1}, \;
S_{5}(Y_{k})=-5 c_{0}.
\label{(B1)}
\end{equation}
On the other hand, we can evaluate $S_{1}(Y_{k})$ and $S_{2}(Y_{k})$ by using the quadratic
transformation (A6) and exploiting the identities
\begin{equation}
S_{1}(Y_{k})=S_{2}(X_{k})+\mu S_{1}(X_{k})+5 \nu,
\label{(B2)}
\end{equation}
\begin{equation}
S_{2}(Y_{k})=S_{4}(X_{k})+2 \mu S_{3}(X_{k})+(\mu^{2}+2 \nu)S_{2}(X_{k})
+2 \mu \nu S_{1}(X_{k})+5 \nu^{2},
\label{(B3)}
\end{equation} 
obtaining therefore the following equations for $\mu$ and $\nu$:
\begin{equation}
\mu a_{4}-5 \nu +2 a_{3}-(a_{4})^{2}=0,
\label{(B4)}
\end{equation}
\begin{equation}
\mu^{2}a_{3}-10 \nu^{2}+\mu (3 a_{2}-a_{3}a_{4})+2a_{1}-2a_{2}a_{4}+(a_{3})^{2}=0,
\label{(B5)}
\end{equation}
where, in the course of arriving at Eq. (B5), we have re-expressed repeatedly $(a_{4})^{2}$ from Eq. (B4).
This system is quadratic with respect to $\mu$ and $\nu$, and hence leads to two sets of coefficients. 
For the case studied in Eq. (2.6), they reduce to (here $a_{3}=\rho_{3},a_{4}=\rho_{4}$)
\begin{equation}
\mu_{\pm}={a_{4}[13a_{3}-4 (a_{4})^{2}] \pm \sqrt{60 (a_{3})^{3}-15 (a_{3}a_{4})^{2}} \over
2 [5a_{3}-2(a_{4})^{2}]},
\label{(B6)}
\end{equation}
\begin{equation}
\nu_{\pm}={\mu_{\pm}\over 5}a_{4}+{2 \over 5}a_{3}-{1 \over 5}(a_{4})^{2}.
\label{(B7)}
\end{equation}
There is complete freedom to choose either of these. After finding $\mu$ and $\nu$ in such a way, one 
can use the Eqs. (B1) to obtain $c_{0},c_{1},c_{2}$. One finds explicitly, in general,
\begin{eqnarray}
c_{0}&=& -\nu^{5}-\mu \nu^{4}S_{1}(X_{k})-(2 \mu^{2}\nu^{3}+\nu^{4})S_{2}(X_{k})
- \left(2\mu^{3}\nu^{2}+4\mu \nu^{3}\right)S_{3}(X_{k})
\nonumber \\
&-& \left(\mu^{4}\nu +6\mu^{2}\nu^{2}+2\nu^{3}\right)S_{4}(X_{k})
-\left({\mu^{5}\over 5}+4\mu^{3}\nu+6 \mu \nu^{2}\right)S_{5}(X_{k})
\nonumber \\
&-& (\mu^{4}+6 \mu^{2}\nu+2 \nu^{2})S_{6}(X_{k})
-(2\mu^{3}+4 \mu \nu) S_{7}(X_{k})-(2 \mu^{2}+\nu)S_{8}(X_{k})
\nonumber \\
&-& \mu S_{9}(X_{k})-{1 \over 5}S_{10}(X_{k}),
\label{(B8)}
\end{eqnarray}
\begin{eqnarray}
c_{1}&=& -{5 \over 4}\nu^{4}-\mu \nu^{3}S_{1}(X_{k})
-\left({3 \over 2}\mu^{2}\nu^{2}+\nu^{3}\right)S_{2}(X_{k})
\nonumber \\
&-& (\mu^{3}\nu+3 \mu \nu^{2})S_{3}(X_{k})
-\left({\mu^{2}\over 4}+3 \mu^{2}\nu+{3 \over 2}\nu^{2}\right)S_{4}(X_{k})
-(\mu^{3}+3 \mu \nu)S_{5}(X_{k})
\nonumber \\
&-& \left({3 \over 2}\mu^{2}+\nu \right)S_{6}(X_{k})
-\mu S_{7}(X_{k})-{1 \over 4}S_{8}(X_{k}),
\label{(B9)}
\end{eqnarray}
\begin{eqnarray}
c_{2}&=& -{5 \over 3}\nu^{3}-\mu \nu^{2}S_{1}(X_{k})-(\mu^{2}\nu+\nu^{2})S_{2}(X_{k})
-{\mu \over 3}(\mu^{2}+6 \nu)S_{3}(X_{k})
\nonumber \\
&-& (\mu^{2}+\nu)S_{4}(X_{k})-\mu S_{5}(X_{k})-{1 \over 3}S_{6}(X_{k}).
\label{(B10)}
\end{eqnarray}

The removal from a general quintic of the three terms in $X^{4},X^{3}$ and $X^{2}$ brings it to the
Bring-Jerrard form (2.10), here re-written for convenience as
\begin{equation}
X^{5}+d_{1}X+d_{0}=0.
\label{(B11)}
\end{equation}
By virtue of the Newton formulas (A3), the power sums for the quintic (B11) are
\begin{equation}
S_{1}(Z_{k})=S_{2}(Z_{k})=S_{3}(Z_{k})=0, \; 
S_{4}(Z_{k})=-4d_{1}, \; 
S_{5}(Z_{k})=-5 d_{0}.
\label{(B12)}
\end{equation}
Assuming now, following Bring \cite{Bring}, that the roots $Z_{k}$ of Eq. (B11) are related by the quartic
transformation (A8) to the roots $Y_{k}$ of the principal quintic (A7), 
we can substitute Eq. (A8) into Eq. (B12). This leads to a system of five equations with six
unknown variables. More precisely, from the equation
\begin{equation}
S_{1}(Z_{k})=5 u_{4}-4c_{1}-3 u_{1}c_{2}=0,
\label{(B13)}
\end{equation}
one finds 
\begin{equation}
u_{4}={4\over 5}c_{1}+{3 \over 5}c_{2}u_{1}.
\label{(B14)}
\end{equation} 
The second equation \cite{Adamchik}
\begin{eqnarray}
S_{2}(Z_{k})&=& -10 u_{1}u_{2}c_{0}-4 (u_{2})^{2}c_{1}
+{4 \over 5}(c_{1})^{2}+8 c_{0}c_{2}+{46 \over 5}u_{1}c_{1}c_{2} 
\nonumber \\
&+& \left[{6 \over 5}(u_{1})^{2}+6 u_{2}\right](c_{2})^{2}
-2 u_{3}(5c_{0}+4 u_{1}c_{1}+3 u_{2}c_{2})=0,
\label{(B15)}
\end{eqnarray}
obtained from the identities
\begin{eqnarray}
S_{2}(Z_{k})&=& S_{8}(Y_{k})+2 u_{1}S_{7}(Y_{k})+[(u_{1})^{2}+2 u_{2}]S_{6}(Y_{k})
+2(u_{1}u_{2}+u_{3})S_{5}(Y_{k}) 
\nonumber \\
&+& [(u_{2})^{2}+2 u_{4}+2 u_{1}u_{3}]S_{4}(Y_{k})
+2(u_{2}u_{3}+u_{1}u_{4})S_{3}(Y_{k})+5(u_{4})^{2},
\label{(B16)}
\end{eqnarray}
\begin{equation}
S_{6}(Y_{k})=3(c_{2})^{2}, \;
S_{7}(Y_{k})=7 c_{1}c_{2}, \;
S_{8}(Y_{k})=8 c_{0}c_{2}+4 (c_{1})^{2},
\label{(B17)}
\end{equation}
relates $u_{2}$ and $u_{3}$. The clever idea of the Bring-Jerrard method lies in choosing
$u_{2}$ in such a way that the coefficient of $u_{3}$ in Eq. (B15) vanishes. By inspection
one finds immediately
\begin{equation}
u_{2}=-{5 \over 3}{c_{0}\over c_{2}}
-{4 \over 3}{c_{1}\over c_{2}}u_{1}.
\label{(B18)}
\end{equation}
Thus, Eq. (B15) now depends only on $u_{1}$ and is a quadratic, i.e. \cite{Adamchik}
\begin{eqnarray}
\; & \; & \Bigr[27 (c_{2})^{4}-160 (c_{1})^{3}+300 c_{0}c_{1}c_{2} \Bigr](u_{1})^{2}
+\Bigr[27 c_{1}(c_{2})^{3}-400 c_{0}(c_{1})^{2}
+375 (c_{0})^{2}c_{2} \Bigr]u_{1} 
\nonumber \\
&+& 18 (c_{1}c_{2})^{2}-45c_{0}(c_{2})^{3}-250 (c_{0})^{2}c_{1}=0.
\label{(B19)}
\end{eqnarray}
Last, by setting the sum of the cubes of (A8) to zero by virtue of (B12), a cubic equation 
for $u_{3}$ is obtained, by virtue of the identity
\begin{equation}
S_{3}(Z_{k})=5(u_{4})^{3}+\sum_{l=2}^{12}b_{l}S_{l}(Y_{k}),
\label{(B20)}
\end{equation}
where (recall that we already know $S_{1}(Y_{k})...S_{8}(Y_{k})$)
\begin{equation}
b_{2}=3 u_{2}(u_{4})^{2},
\label{(B21)}
\end{equation}
\begin{equation}
b_{3}=(u_{3})^{3}+3 u_{1}(u_{4})^{2}+6 u_{2}u_{3}u_{4},
\label{(B22)}
\end{equation}
\begin{equation}
b_{4}=3(u_{2})^{2}u_{4}+3(u_{4})^{2}+3 u_{2}(u_{3})^{2}
+6 u_{1}u_{3}u_{4},
\label{(B23)}
\end{equation}
\begin{equation}
b_{5}=3(u_{2})^{2}u_{3}+3 u_{1}(u_{3})^{2}
+6u_{4}(u_{3}+u_{1}u_{2}),
\label{(B24)}
\end{equation}
\begin{equation}
b_{6}=(u_{2})^{3}+3(u_{1})^{2}u_{4}+3(u_{3})^{2}
+6 u_{2}(u_{4}+u_{1}u_{3}),
\label{(B25)}
\end{equation}
\begin{equation}
b_{7}=3(u_{1})^{2} u_{3}+3(u_{2})^{2}u_{1}
+6(u_{1}u_{4}+u_{2}u_{3}),
\label{(B26)}
\end{equation}
\begin{equation}
b_{8}=3 u_{4}+3(u_{1})^{2}u_{2}+3(u_{2})^{2}
+6 u_{1}u_{3},
\label{(B27)}
\end{equation}
\begin{equation}
b_{9}=(u_{1})^{3}+3 u_{3}+6 u_{1}u_{2},
\label{(B28)}
\end{equation}
\begin{equation}
b_{10}=3 u_{2}+3(u_{1})^{2},
\label{(B29)}
\end{equation}
\begin{equation}
b_{11}=3 u_{1},
\label{(B30)}
\end{equation}
\begin{equation}
b_{12}=1,
\label{(B31)}
\end{equation}
\begin{equation}
S_{9}(Y_{k})=9c_{0}c_{1}-3(c_{2})^{3},
\label{(B32)}
\end{equation}
\begin{equation}
S_{10}(Y_{k})=5(c_{0})^{2}-10c_{1}(c_{2})^{2},
\label{(B33)}
\end{equation}
\begin{equation}
S_{11}(Y_{k})=-11c_{0}(c_{2})^{2}-11(c_{1})^{2}c_{2},
\label{(B34)}
\end{equation}
\begin{equation}
S_{12}(Y_{k})=-24c_{0}c_{1}c_{2}-4(c_{1})^{3}+3(c_{2})^{4}.
\label{(B35)}
\end{equation}
All intermediate quantities for reduction to the Bring-Jerrard form
can be therefore found in terms of radicals. 
Of course, once the irreducible quintic (B11) is solved in terms of
hypergeometric functions as outlined in Sec. II, one has to invert the Tschirnaus-Bring quartic 
transformation (A8) to obtain the solutions of the principal 
quintic (A7) and, eventually, of the original quintic (A1), i.e.,
\begin{equation}
X_{k}=X_{k}(Y_{j}(Z_{l})).
\label{(B36)}
\end{equation} 
Thus, one obtains in general twenty candidates for five solutions, and only
numerical testing can tell which ones are correct \cite{Adamchik}.

\end{appendix}

\end{document}